\documentclass[sigconf]{acmart}

\AtBeginDocument{%
  \providecommand\BibTeX{{%
    Bib\TeX}}}
\setcopyright{acmlicensed}
\copyrightyear{2026}
\acmYear{2026}
\acmDOI{XX.XX}
\acmConference[]{MobiHoc '26}{November 23--26,
  2026}{Tokyo, Japan}
\acmISBN{978-1-4503-XXXX-X/2018/06}

\usepackage{algorithm} 
\usepackage{algorithmic}
\usepackage{graphicx}
\usepackage{textcomp}
\usepackage{xcolor}
\def\BibTeX{{\rm B\kern-.05em{\sc i\kern-.025em b}\kern-.08em
    T\kern-.1667em\lower.7ex\hbox{E}\kern-.125emX}}
\usepackage{float}
\usepackage{url} 
\usepackage{subcaption}
\usepackage{multirow}
\usepackage{array}
\usepackage{tikz}
\usepackage[table]{xcolor}
\usepackage{booktabs}
\usepackage{wrapfig}
\newcommand*\blkcirc[1]{\tikz[baseline=(char.base)]{
            \node[shape=circle,fill,inner sep=0.6pt] (char) {\textcolor{white}{#1}};}}

\newcommand{\bva}{\mathbf{v_a}}

\newcommand{\bz}{\mathbf{z}}
\newtheorem{assumption}{Assumption}
\newtheorem{proposition}{Proposition}

\begin{document}

\title{MEO: Mining Reliable Expert Signals for Offline Reinforcement Learning in Wireless Networks}

\author{Lipeng Zu$^{1}$, Hansong Zhou$^{1}$, Yu Qian$^{1}$, Shayok Chakraborty$^{1}$, Yukun Yuan$^{2}$, \\Linke Guo$^{3}$, Xiaonan Zhang$^{1}$}

\affiliation{
  \institution{$^{1}$Florida State University, Florida, USA \\ 
        $^{2}$University of Tennessee at Chattanooga, Tennessee, USA \\
        $^{3}$Clemson University, South Carolina, USA}
  \city{}
  \state{}
  \country{}
}
\email{{lz23b, hz21e, yq23a, schakraborty2, xzhang14}@fsu.edu, yukun-yuan@utc.edu, linkeg@clemson.edu}







\renewcommand{\shortauthors}{Zu et al.}

\begin{abstract}
The operation of next-generation wireless networks increasingly hinges on Deep Reinforcement Learning (DRL) to optimize critical network decisions. Offline RL trains agents using accessible datasets and thus eliminates costly online interactions, rendering it particularly advantageous in dynamic wireless environments that demand low latency and high safety. However, its capability remains underexplored in wireless settings, where offline data are often limited and subject to partial observability. As a result, the learned policy fails to fully capture the informative decision patterns hidden in the dataset. To bridge this gap, we propose MEO, a framework that mines reliable expert signals from wireless datasets for offline RL deployment. MEO first identifies reliable expert samples from imperfect offline data, and then learns a structured latent representation to better distinguish expert and non-expert behaviors. Based on the representation, MEO further derives a compensable reward signal that helps offline RL algorithms more effectively exploit valuable expert knowledge. We evaluate MEO in three practical wireless network scenarios using datasets collected from real-world deployments. The results show that MEO consistently outperforms existing methods across all three scenarios. Notably, MEO can be seamlessly integrated with any offline RL algorithms, highlighting its generalizability and potential for wide adoption in wireless networks. \textbf{(Code is available at \url{https://github.com/lpzu/MEO})}
 
\end{abstract}




\maketitle

\section{Introduction}
\subsection{Background: RL in Wireless Networks}
Next-generation wireless networks are expected to support significantly higher data rates, lower latency, and a massive number of connected devices~\citep{kiruthiga2025evolution}. Meeting these requirements poses significant challenges for network control, as wireless environments are becoming increasingly dynamic and difficult to model. Traditional rule-based and optimization-driven approaches  fall short under such conditions, due to their reliance on static models and fixed heuristics~\cite{Nguyen2021AINetwork}. In contrast, learning-based methods offer greater flexibility by adapting to changing environments and extracting actionable patterns from data. Among them, reinforcement learning (RL) has shown strong potential for sequential decision-making, with demonstrated success in tasks such as spectrum allocation~\cite{yang2024offline}, network scheduling~\cite{bonati2022openrangym}, and edge offloading~\cite{Tan2022offloading}. 

Despite its promise, existing RL approaches in wireless networks rely on online interaction with the environment, where the agent continuously updates its policy through trial-and-error learning~\cite{Yang2025OnlineRL}. In dynamic and safety-critical wireless environment, such as vehicular networks, UAV-assisted communications, industrial IoT systems, and mission-critical public safety networks, online exploration can cause unstable behavior, degraded the quality of service, and even service disruptions~\cite{Yang2024OnlineIss}. In contrast, offline RL learns solely from pre-collected datasets without requiring further interaction with the environment. By eliminating unsafe exploration during online execution, it substantially reduces operational risks, and thus is more suitable for real-world wireless networks~\cite{Fig2024OfflineSurv}. 

Nevertheless, offline RL remains underexplored in wireless networks. An important reason is that offline RL depends entirely on pre-collected dataset, while wireless data collection faces fundamental barriers. On one hand, wireless networks are often subject to partial observability. For example, in handoff management for mobile users, the controller observes the current large-scale fading experienced by the user, but cannot directly access the future channel evolution induced by user mobility~\citep{ammar2024handoffs}. As a result, the collected dataset provides only a partial and potentially biased view of the environment. On the other hand, collecting sufficient dataset is challenging due to the need for long-term system operation and the presence of privacy concerns~\citep{oliveira2021generating}. Consequently, the available dataset is limited and imperfect. Therefore, a key question arises: \textit{how to learn reliable policies from limited and biased offline data?}

\subsection{State of the Art and Challenges} 
A common approach to improving the reliability of offline RL is to leverage high-quality samples that approximate expert behaviors, typically generated by human experts or high-performing policies~\cite{levine2020offline}. However, expert behavioral samples are even scarcer due to limited data collection~\cite{qiao2023privacy, hippalgaonkar2023knowledge}. As a more practical alternative, recent work explores \textit{reward relabeling}, which reconstructs the reward structure across the dataset by propagating guidance from a small number of expert-labeled states and actions~\cite{yan2024simple}. To enable robust offline policy learning in wireless networks, we identify and address two key challenges:

\textbf{Challenge}~\blkcirc{1} \textbf{: How to identify expert behavioral samples within the dataset under partial observability.}
Reward relabeling begins by selecting the highest-return trajectories as expert samples within the dataset. However, this strategy can be biased in wireless networks due to partial observability. Hidden environmental factors can influence the reward without being reflected in the observable state, causing trajectories with high returns to arise from favorable regimes rather than better decisions. For example, in task offloading for multi-access edge computing networks, the highest Quality-of-Service (QoS) trajectories often occur during periods of low server load and stable network connectivity. Treating them as expert data can bias learning and reduce robustness under dynamic network conditions.

\textbf{Challenge}~\blkcirc{2} \textbf{: How to learn a representation space that can effectively exploit identified expert samples.}
In offline RL for wireless networks, expert samples are typically sparse and embedded within a much larger set of heterogeneous non-expert behaviors. To make expert knowledge useful beyond the labeled samples themselves, the learned representation space must preserve behaviorally meaningful structure, so that expert and non-expert data can be compared and organized in a reliable manner. If the representation space fails to reflect such structure, expert information cannot be effectively propagated across the dataset, and its supervisory value becomes severely diluted. Therefore, learning a structure-aware representation space that supports effective utilization of expert samples remains a fundamental challenge.

\subsection{Present Work}
\begin{figure}
\centering
    \includegraphics[width=0.95\linewidth]{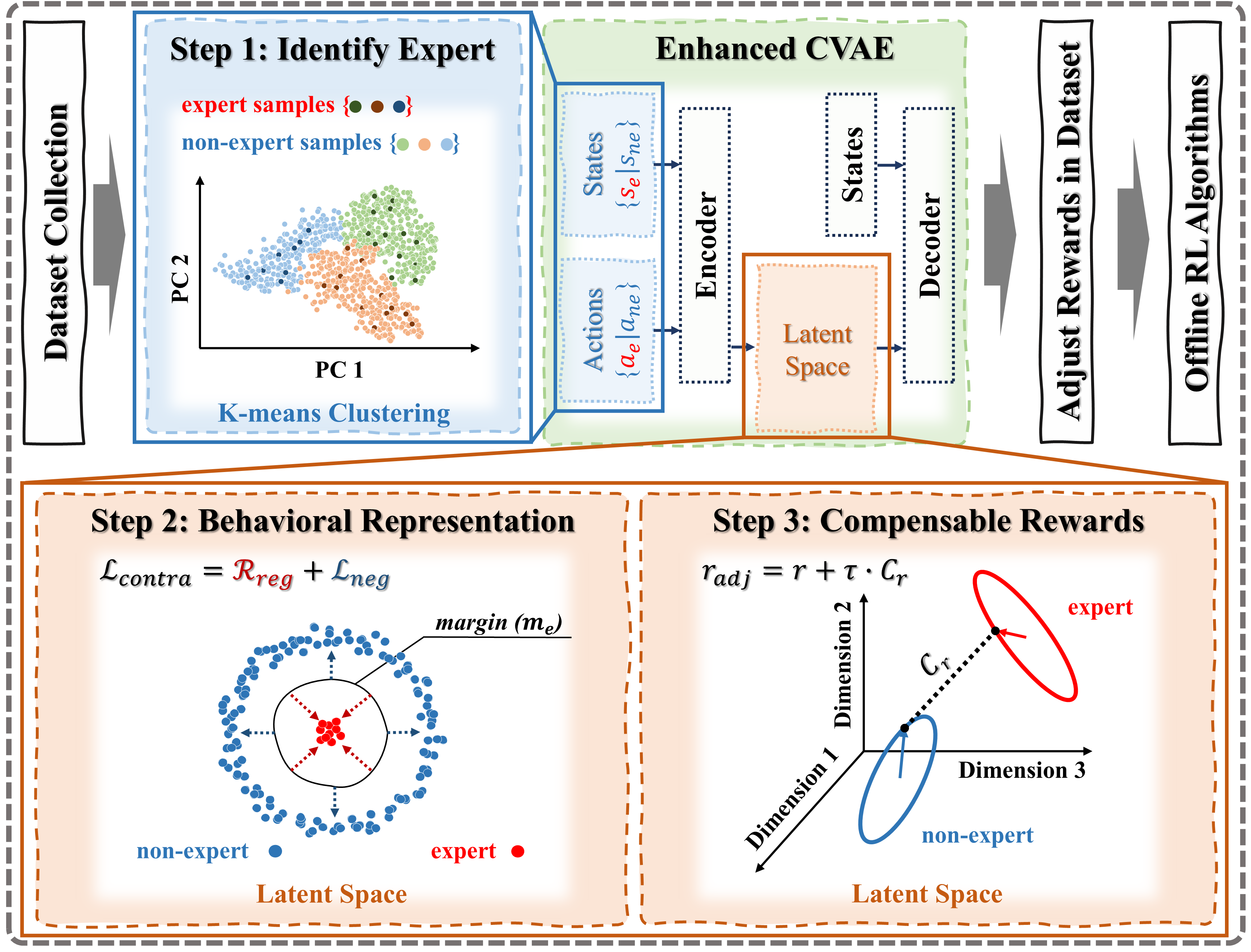}
    \caption{Overview of our proposed framework MEO.}
    \Description{}
    \label{fig:framework}
\end{figure}

We propose MEO, an general offline RL framework mining reliable expert signals from collected datasets in wireless networks, as illustrated in Fig.~\ref{fig:framework}. To address challenge \blkcirc{1}, \textbf{Step 1} (Sec.~\ref{Sec: Step 1}) identifies expert samples from the offline dataset using a cluster-wise Top-$K$ strategy, which provides a more reliable identification of expert samples by mitigating hidden-state confounding under partial observability. To address challenge \blkcirc{2}, \textbf{Step 2} (Sec.~\ref{Sec: Step 2}) incorporates the contrastive regularization  into the Conditional Variational Autoencoder (CVAE) objective to improve representation disentanglement and enhance latent discriminability. In \textbf{Step 3} (Sec.~\ref{Sec: Step 3}), MEO measures the latent distance between each sample and the expert centroid and uses it as a compensable reward to reflect behavioral proximity to the expert behaviors, improving the utilization of expert information during reward relabeling.

We validate MEO with three representative wireless network use cases in real world. In the first one, we investigate the resource allocation in ORAN-Cloud with multiple available servers deployed within the centralized unit. The agent distributes incoming tasks to appropriate servers based on user demand, hardware status and task queue to maximize the system-wide QoS. The second one is to offload real-time visual tasks from edge devices (NVIDIA Orin Nano) to server (NVIDIA RTX 3060 Ti) for inference acceleration. The agent needs to balance latency and energy consumption by adjusting the offloading portions based on the system status. In the last one, we set up 2 USRP N210s to implement CSI-based (Channel State Information) adaptive random channel access for data transmission using Wi-Fi protocol in dynamic network conditions. 

We summarize our key contributions as follows.
\begin{itemize}
    \item We analyze the limitations of trajectory-level expert selection under partial observability and propose a cluster-wise Top-$K$ method that identifies expert behaviors.

    \item We develop a contrastive-enhanced CVAE, which produces discriminative behavioral representations under stochastic and context-dependent network dynamics.

    \item We convert latent proximity to expert behaviors into a compensable reward signal, improving reward variation and offline learning efficiency from imperfect data.

    \item We evaluate MEO on the datasets collected from multiple real-world wireless systems, where it consistently improves policy reliability and robustness while remaining compatible with existing offline RL algorithms.
\end{itemize}

\section{Preliminaries}
\subsection{POMDP Formulation}
A Partially Observable Markov Decision Process (POMDP) extends a Markov Decision Process (MDP) to the setting where the full environment state cannot be directly observed. Formally, a POMDP is defined by the tuple \((\mathcal{S}, \mathcal{H}, \mathcal{A}, \mathcal{T}, r, \rho_0, \gamma)\), where \(\mathcal{S}\) denotes the observable state space, \(\mathcal{H}\) denotes the hidden state space, and \(\mathcal{A}\) denotes the action space. The transition \(\mathcal{T}\) specifies the dynamics of the observable and hidden states conditioned on the state-action pair, the reward function is \(r: \mathcal{S} \times \mathcal{H} \times \mathcal{A} \rightarrow \mathbb{R}\), \(\rho_0\) denotes the initial state distribution, and \(\gamma \in [0,1)\) is the discount factor.

\subsection{Hidden State in a Causal Graph}
\begin{figure}[H]
\centering
\includegraphics[width=0.7\linewidth]{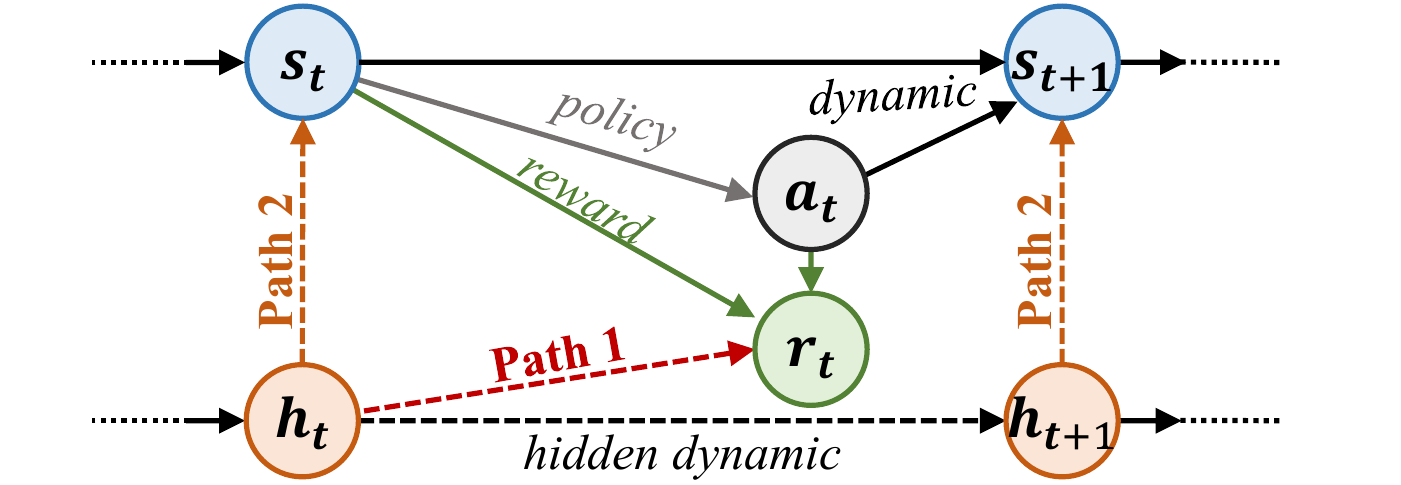}
\caption{Causal graph of the POMDP.}
\label{fig:causal_graph}
\end{figure}

A causal graph offers a principled way to characterize how observable variables and latent factors jointly influence the decision process. Following the causal modeling under POMDP~\cite{pace2023delphic}, the causal graph in Fig.~\ref{fig:causal_graph} is adopted to characterize the effect of the hidden state. In particular, the hidden state $h_t \in \mathcal{H}$ affects the reward $r_t$ through two pathways: a direct \textit{Path 1} from $h_t$ to $r_t$, and an indirect \textit{Path 2} mediated by the observable state $s_t$ and action $a_t$.

\subsection{Conditional VAE}
Conditional Variational Autoencoder (CVAE) is a generative model that extends the Variational Autoencoder (VAE) by incorporating conditional variables. Mathematically, CVAE aims to maximize the Evidence Lower Bound (ELBO) on the conditional log-likelihood of observing an action given a state~\cite{sohn2015learning}, formulated as follows 
\begin{equation}\label{eqn: original CVAE loss}
\begin{aligned}
    \mathcal{L} & _{\text{CVAE}}(s, a; \theta, \phi) = \mathcal{L} _{\text{recon}} + \mathcal{L}_{\text{KL}} \\
    &= -\mathbb{E}_{q_\phi(\bz|s, a)} \left[ \log p_\theta(a|\bz, s) \right] + \text{KL}\left(q_\phi(\bz|s, a) \Vert p_\theta(\bz|s)\right),
\end{aligned}
\end{equation}
where: $\mathcal{L} _{\text{recon}}$ and KL indicate the reconstruction loss and KL-divergence, respectively; \( \bz \) is the latent variable encoded by the posterior \( q_\phi(\bz|s,\bva) \); \( p_\theta(a|\bz, s) \) is the likelihood of reconstructing the action given the latent variables and the state; and \( p_\theta(\bz|s) \) represents the prior. 

\section{Method}\label{sec:method}
We propose \textbf{MEO}, a framework that enables reliable offline RL for wireless networks under partial observability. Specifically, MEO first identifies high-quality expert samples from the offline dataset to focus learning on informative decision patterns (Sec.~\ref{Sec: Step 1}). Behavior-aware representations are then learned to better separate expert and non-expert decisions (Sec.~\ref{Sec: Step 2}). In the end, MEO incorporates the expert signals from the representation space into reward relabeling for policy optimization (Sec.~\ref{Sec: Step 3}).

\subsection{Expert Data Selection}\label{Sec: Step 1}
\subsubsection{Limitations of Maximum-Return Trajectory}
A common strategy for identifying expert data is to treat trajectories with the highest observed returns as expert trajectories. However, this strategy can be biased in wireless networks, which are often both highly dynamic and partially observable. For example, in edge offloading, the agent may observe current queue lengths, server utilization, and channel measurements, but cannot fully observe future channel variations, background interference, or workload evolution.

As illustrated in Fig.~\ref{fig:causal_graph}, the reward depends not only on the observable state-action pair $(s_t,a_t)$, but also on hidden state $h_t$. Therefore, even when the observable state and action are same, the realized reward is likely to vary across different hidden states, i.e.,
\begin{equation}
    \mathrm{Var}(r_t \mid s_t=s, a_t=a) > 0.
\end{equation}

Moreover, since the trajectory return is the discounted accumulation of per-step rewards,
\begin{equation}
    G = \sum_{t=0}^{T-1} \gamma^t r_t
      = \sum_{t=0}^{T-1} \gamma^t r(s_t,a_t,h_t),
\end{equation}
it is likewise affected by hidden environmental factors over time. A trajectory with a high observed return may therefore reflect a favorable hidden regime rather than better decision-making, so maximum-return selection fails to identify truly expert behavior.

\subsubsection{Cluster-wise Top-$K$ Sample Selection}
To address the above limitation, we move from trajectory-based selection to sample-based selection. From the causal graph in Fig.~\ref{fig:causal_graph}, the hidden state $h$ affects the reward through an indirect path $\eta(s,a;h)$ mediated by the observable state and action, and a direct path $\delta(h)$, so that $r(s,a,h) = g(\eta(s,a;h),\delta(h))$ with $g$ combining the two effects. We then impose the following regularity conditions.
\begin{assumption}\label{asm:1}
The conditional hidden-state distribution is locally stable in the observable state, i.e.
\[
  W_1\big(P(h\mid s_i),\,P(h\mid s_j)\big)\le L_{\mathcal{H}}\|s_i-s_j\|
\]
where $W_1$ denotes the first-order Wasserstein distance, and the indirect effect is uniformly $L_\eta$-Lipschitz in the hidden state,
\(
  |\eta(s,a;h)-\eta(s,a;h')|\le L_\eta\, d_{\mathcal{H}}(h,h')
\)
with $d_{\mathcal{H}}$ a metric on $\mathcal{H}$.
\end{assumption}
\begin{proposition}\label{prop:1}
Under the causal view in Fig.~\ref{fig:causal_graph} and Assumption~\ref{asm:1}, cluster-wise Top-$K$ selection reduces the dependence of reward-based sample ranking on the hidden state $h_t$.
\end{proposition}
\begin{proof}
Cluster-wise Top-$K$ groups samples with similar observable states, so that any two samples $u_i=(s_i,a_i)$ and $u_j=(s_j,a_j)$ in the same cluster satisfy $\|s_i - s_j\| \le \varepsilon_s$. Let $\bar\eta_{s}(u)=\mathbb{E}_{h\sim P(h\mid s)}[\eta(u;h)]$ denote the indirect effect of a state-action pair $u$ evaluated under the hidden-state distribution induced by $s$, where $s$ need not be the observable state of $u$. The difference between the two samples then splits as
\[
\bar\eta_{s_i}(u_i)-\bar\eta_{s_j}(u_j)
=\underbrace{\bar\eta_{s_i}(u_i)-\bar\eta_{s_j}(u_i)}_{\text{hidden-state discrepancy}}
+\underbrace{\bar\eta_{s_j}(u_i)-\bar\eta_{s_j}(u_j)}_{\text{common distribution}},
\]
where the second term compares the two state-action pairs under a single hidden-state distribution and is therefore free of hidden-state influence. For the first term, $\eta(u;\cdot)$ is $L_\eta$-Lipschitz in $h$ by Assumption~\ref{asm:1}, so the Kantorovich--Rubinstein duality gives
\[
\big|\bar\eta_{s_i}(u_i)-\bar\eta_{s_j}(u_i)\big|
\le L_\eta\, W_1\big(P(h\mid s_i),P(h\mid s_j)\big)
\le L_\eta L_{\mathcal{H}}\,\varepsilon_s .
\]
The hidden-state discrepancy along the indirect path thus shrinks linearly with the cluster diameter, so that within a cluster the indirect effect is determined by the comparison under a common distribution up to an error of at most $L_\eta L_{\mathcal{H}}\varepsilon_s$.
\end{proof}
Clustering can further reduce the within-cluster variation of $\delta(h)$ when observable states reflect hidden regimes, which is plausible in wireless systems, where clusters with long queues and high utilization likely share similar latent workloads. 

Based on Prop.~\ref{prop:1}, we design a cluster-wise Top-$K$ expert selection strategy to identify reliable expert samples from the offline dataset. Specifically, we first apply K-means clustering over the observable states. Then, within each cluster, samples are ranked according to their rewards, and the top-$K$ samples are selected as experts. These selected samples form the expert dataset $\mathcal{D}^e$, while the remaining samples are assigned to the non-expert dataset $\mathcal{D}^{ne}$. We refer to the samples in $\mathcal{D}^e$ as \emph{pseudo-experts}, since they are the highest-reward samples within clusters of similar observable states, not samples drawn from an optimal policy. For brevity we continue to write $\mathcal{D}^e$ and to speak of expert samples in this local sense.

\subsection{Expert-Guided Latent Space Structuring}\label{Sec: Step 2}
\subsubsection{Latent Modeling and Expert Regularization}
Given the identified expert samples, the next critical challenge is to enable effective exploitation of expert information. While direct comparison in the original state--action space is often unreliable, proximity in raw features does not necessarily reflect similarity in decision behavior under heterogeneous states and actions. This motivates the need for a transformed representation space where samples can be compared more meaningfully. To this end, we learn a behavior-aware latent representation by adopting a CVAE to encode state--action pairs, as it models actions conditioned on states and provides a structured space for subsequent expert-guided comparison.

\begin{figure}[H]
    \centering
    \includegraphics[width=0.98\linewidth]{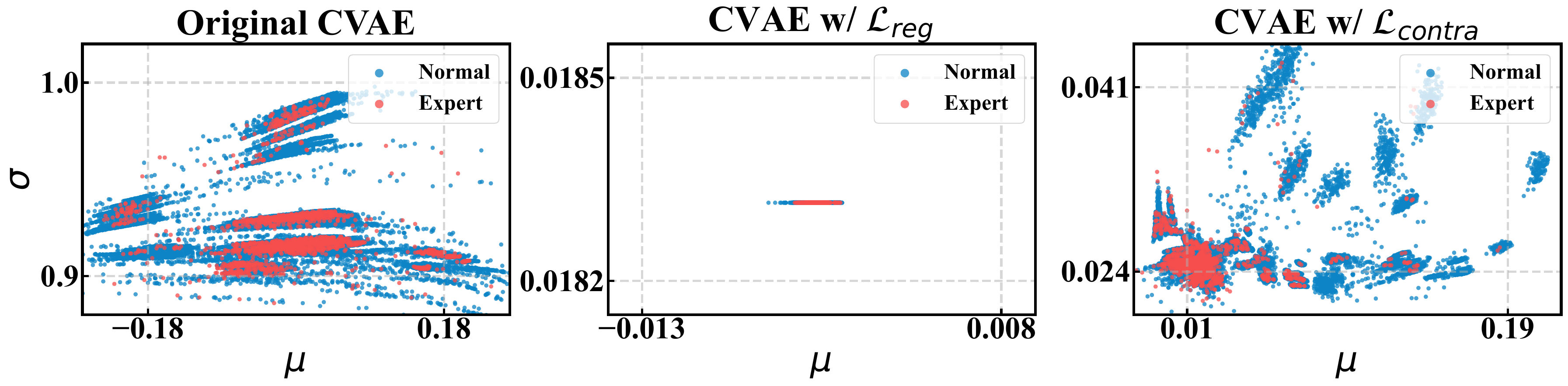}
    \caption{Visualization of latent representations on the edge computing dataset. (\textit{left}) Standard CVAE. (\textit{middle}) Expert regularization alone. (\textit{right}) Adding contrastive loss.}
    \Description{}
    \label{fig:latent}
\end{figure}

However, conventional CVAE training is primarily driven by reconstruction and distribution matching (See Eq.~\ref{eqn: original CVAE loss}), which does not explicitly encourage separation between expert and non-expert samples. Consequently, the posterior distributions of expert and non-expert samples in the latent space still overlap substantially, as illustrated in Fig.~\ref{fig:latent}(\textit{left}). Here, each point corresponds to the posterior parameters $(\mu, \sigma)$ produced by the CVAE encoder for a state--action sample, where $\mu$ and $\sigma$ denote the mean and standard deviation of the latent Gaussian distribution $q_\phi(z \mid s,a)$.

To better organize expert behaviors in the latent space, a natural approach is to encourage expert samples to form a compact cluster by penalizing the dispersion of their representations \cite{liu2023clue}
\begin{equation}\label{Eq:regLoss}
    \mathcal{L}_{\text{reg}} = \lambda_{\text{reg}} \cdot (\text{mean}(\mathbf{z}_e)+\text{std}(\mathbf{z}_e)),
\end{equation}
where \(\lambda_{\text{reg}}\) controls the strength of the regularization.
This regularization provides a useful starting point, since it encourages expert samples to occupy a concentrated region in the latent space. However, it mainly promotes compactness of expert embeddings, without explicitly enforcing separation between expert and non-expert behaviors. As a result, when the two groups exhibit similar state--action patterns, the encoder may satisfy the regularization objective by excessively compressing all representations into a narrow region (almost compress to a point), as illustrated in Fig.~\ref{fig:latent}(\textit{middle}).

\subsubsection{Enhancing CVAE with Contrastive Separation}
To achieve both compact expert representations and explicit separation from non-expert behaviors, we enhance the CVAE objective with a margin-based contrastive regularizer. Since \(\mathcal{L}_{\text{reg}}\) already encourages compact expert embeddings, the additional objective focuses on pushing expert and non-expert samples apart in the latent space. Specifically, we define a negative-pair loss
\begin{equation}\label{Eq:L_neg}
    \mathcal{L}_{\text{neg}} =
    \lambda_{\text{neg}}
    \mathbb{E}_{e \sim \mathcal{D}^e,\, ne \sim \mathcal{D}^{ne}}
    \left[
        \max\left(0, m_e - \|\mathbf{z}_e - \mathbf{z}_{ne}\|_2 \right)
    \right],
\end{equation}
where \(\mathbf{z}_e\) and \(\mathbf{z}_{ne}\) denote the latent representations of expert and non-expert samples, \(m_e\) is a margin enforcing a minimum separation between them, and \(\lambda_{\text{neg}}\) controls the strength of the loss.

Together with the regularization \(\mathcal{L}_{\text{reg}}\) in Eq.~\ref{Eq:regLoss}, the negative-pair loss forms the contrastive objective \(\mathcal{L}_{\text{contra}}\), and the overall objective of the enhanced CVAE becomes
\begin{equation}\label{Eq:all}
  \widetilde{\mathcal{L}}_{\text{CVAE}} = \mathcal{L}_{\text{ELBO}} + \mathcal{L}_{\text{contra}},
  \quad \text{with } \mathcal{L}_{\text{contra}} = \mathcal{L}_{\text{reg}} + \mathcal{L}_{\text{neg}}
\end{equation}
The proposed objective encourages the latent space to satisfy two desirable properties: (i) expert samples remain compact, and (ii) expert and non-expert samples are separated by a sufficient margin. As illustrated in Fig.~\ref{fig:latent}(\textit{right}), the resulting latent space exhibits improved separation between expert and non-expert samples while preserving compact expert representations.

\subsection{Compensable Reward}\label{Sec: Step 3}
\subsubsection{Why Expert Data Is More Informative}
While we have already learned a latent representation space that distinguishes expert and non-expert behaviors, a key question remains: how can expert information be exploited to improve the learning signal in offline RL? Empirically, datasets generated by expert behaviors often lead to better training performance, suggesting that expert data carries additional information beyond the raw task reward.

To better understand the difference between expert and non-expert datasets, we conduct a comparative analysis across multiple tasks, including Adroit~\cite{Adroit2018}, Gym-MuJoCo~\cite{todorov2012mujoco}, and a wireless networking scenario, ORAN-Cloud. The first two datasets are obtained from the D4RL benchmark~\cite{fu2020d4rl}, while the ORAN-Cloud datasets are generated in our simulation\footnote{The D4RL datasets are available at \url{http://rail.eecs.berkeley.edu/datasets/offline_rl/}. The non-expert datasets in Adroit and Gym-MuJoCo are collected using behavior cloning and random policies, respectively. For the ORAN-Cloud task, the expert dataset is generated using a trained DDQN policy~\cite{van2016deep}, while the non-expert dataset is produced using a Round-Robin heuristic.}. For each dataset, we cluster samples according to their observable states and measure the reward difference between the highest- and lowest-reward samples within each cluster. This metric captures how well the reward distinguishes samples collected under similar observable contexts, and thus serves as a proxy for the informativeness of the learning signal.

\begin{figure}[t]
    \centering
    \includegraphics[width=0.98\linewidth]{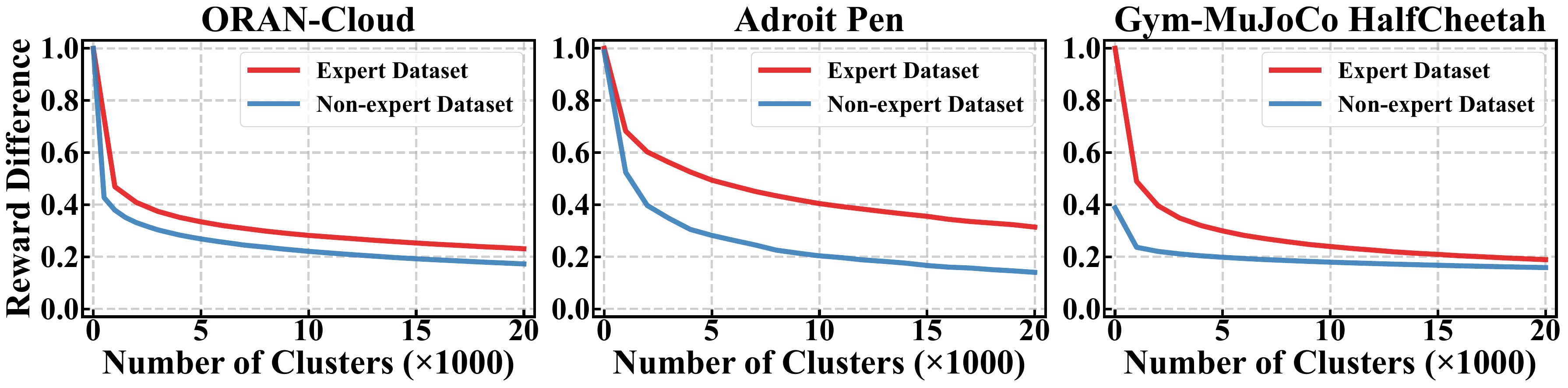}
    \caption{Comparative analysis of expert and non-expert datasets across multiple applications.}
    \Description{}
    \label{fig:motivation}
\end{figure}

As shown in Fig.~\ref{fig:motivation}, expert datasets (red lines) consistently exhibit larger intra-cluster reward differences, indicating stronger local reward discrimination among behaviorally similar samples. In contrast, reward differences in non-expert datasets (blue lines) diminish rapidly as the number of clusters increases, suggesting that non-expert data provides weaker guidance for distinguishing behavior quality within similar observable contexts. Since prior work has shown that reward discrepancy can correlate positively with offline RL performance~\cite{balazadeh2024sequential}, this observation suggests that expert data can provide more informative learning signals for offline policy optimization. Since the datasets in our settings carry no such labels, MEO relies on pseudo-experts to reproduce this property.

\subsubsection{Compensable Reward Mechanism}

Motivated by the above observation, we leverage the learned latent representation to refine the reward signal in the offline dataset. The key idea is that samples closer to expert behaviors in the latent space should receive more favorable training signals, while samples farther from expert behaviors should be penalized. In this way, the reward becomes more behavior-aware and provides additional discrimination beyond the original task reward. 

Specifically, we first compute the centroid of expert samples in the latent space
\begin{equation}\label{Eq:center}
    \mathbf{z}_e^c = \frac{1}{N} \sum_{i=1}^{N} \mathbf{z}_e^i,
\end{equation}
where \( \mathbf{z}_e^i \) denotes the latent representation of the \( i \)-th expert sample in \( \mathcal{D}^e \), and \( N \) is the number of expert samples. The resulting centroid provides a representative latent prototype for expert behavior.

For any sample \(x \in \mathcal{D}\), we quantify its proximity to expert behavior as
\begin{equation}\label{Eq:dist}
    \mathcal{C}_r = - \|\mathbf{z}_x - \mathbf{z}_e^c\|_2 .
\end{equation}

We then relabel the reward in a compensatory manner as
\begin{equation}
    r_{\text{adj}} = r_{\text{ori}} + \tau \cdot \mathcal{C}_r,
    \label{Eq:tau}
\end{equation}
where \( \tau \) controls the strength of the expert-guided compensation and \(r_{\text{ori}}\) denotes the original reward in the dataset. In this way, the original reward is preserved as the primary task signal, while the additional term improves behavioral discrimination.

\subsection{Algorithm Summary}
Algorithm~\ref{algo} summarizes the overall MEO procedure.
\begin{algorithm}[H]
\caption{\textbf{MEO}}
\label{algo}
\begin{algorithmic}[1]
\STATE \textbf{Initialize:} Offline dataset $\mathcal{D}$.
\STATE Cluster-wise Top-$k$ selection on $\mathcal{D}$ to obtain $\mathcal{D}^e$ and $\mathcal{D}^{ne}$.
\STATE Train the CVAE using $\mathcal{D}^e$ and $\mathcal{D}^{ne}$ (Eq.~\ref{Eq:all}).
\STATE Compute the expert centroid in the latent space (Eq.~\ref{Eq:center}).
\STATE Compute the latent distance to the expert centroid (Eq.~\ref{Eq:dist}).
\STATE Adjust rewards according to Eq.~\ref{Eq:tau}.
\STATE Train the offline RL policy with adjusted rewards.
\end{algorithmic}
\end{algorithm}

\section{Deployment to Practical Wireless Networks}
To demonstrate the effectiveness of the MEO framework, we consider three representative wireless use cases for comprehensive empirical evaluation, including scalable and efficient resource allocation in ORAN-Cloud, energy-efficient and low-latency task offloading in edge computing, and CSI-based channel selection under partial observability for Wi-Fi data transmission. These diverse tasks highlight the versatility and robustness of the MEO framework in addressing key challenges in next-generation wireless network management and optimization. 

\subsection{Resource Allocation in ORAN-Cloud}
In an ORAN-Cloud environment, a collection of servers $\mathcal{M}$ deployed within the Centralized Unit (CU) manage various computational tasks, including Radio Resource Management (RRM), Control Plane Processing, Network Slicing, and Machine Learning-based Inference. The CU supports diverse users such as User Equipment (UEs), Network Slices, Distributed Units (DUs), and External Network Services by dynamically allocating resources.  When servers reach maximum capacity, incoming requests are temporarily queued, with adaptive load balancing to enhance efficiency. 

\noindent\textbf{\textit{Action Space:\ \ }}The discrete action space is denoted by \( \mathcal{A} \), with cardinality \( |\mathcal{A}| = 10 \), involves selecting an appropriate server for each incoming task, \(m \in \mathcal{M}\).

\noindent\textbf{\textit{State Representation:\ \ }}The state space \(\mathcal{S} \in \mathbb{R}^{3 + 4 \times |\mathcal{M}|}\) (\( |\mathcal{M}| = 10 \)) encapsulates the following operational metrics:
\begin{itemize}
    \item Demands of the incoming task (user request), including requirement of CPU (\(c_{\text{req}}\)) and RAM (\(r_{\text{req}}\)), as well as the estimated occupation time (\(t_{\text{occ}}\));
    \item The CPU and RAM utilization rates on each server, represented as \({U}_{\text{cpu}} = \{u_{\text{cpu}}^m|m\in \mathcal{M}\}\) and \({U}_{\text{ram}} = \{u_{\text{ram}}^m|m\in \mathcal{M}\}\), which indicate the current resource load;
    \item The length of the pending queue in each server, represented as \(L_\text{queue} = \{l_{\text{queue}}^m|m\in \mathcal{M}\}\), reflecting the count of pending tasks when CPU and RAM are used to handle other tasks;
    \item A queue penalty vector \(P_{\text{queue}} = \{p_{\text{queue}}^m|m\in \mathcal{M}\}\) from each server.  \(P_{\text{queue}}\) quantifies the delay-induced penalty associated with tasks, where $p_{\text{queue}}^m = \sum_i (c_{\text{req}}^i + r_{\text{req}}^i) \cdot t_{\text{occ}}^i$.
\end{itemize}

\noindent\textbf{\textit{Reward Function:\ \ }} We seek to concurrently minimize power consumption and latency for users. The instantaneous power \( P_{\text{power}} \) is defined as in~\cite{liu2017hierarchical}, while the delay penalty \( P_{\text{latency}} \) is computed as a normalized measure based on queuing penalties across all servers. For a set of \( M \) servers, we define:
\begin{align}
    r &= - \left( w_1 \cdot  P_{\text{power}} + w_2 \cdot P_{\text{latency}} \right) \\
        &= - \left( w_1 \cdot \sum_{m=1}^{M} \left( 2u_{\text{cpu}}^m  (u_{\text{cpu}}^m)^{1.4} \right)
           + w_2 \cdot \frac{p_{\text{queue}}^m}{\sum\nolimits_{m \in \mathcal{M}} p^m_{\text{queue}}} \right). 
\end{align}

\noindent\textbf{\textit{Partial Observability:\ \ }}
Although the agent observes the current task demand and aggregated server statistics, the full system state is not directly accessible. For example, the remaining service times of running tasks and the detailed composition of queued jobs are hidden from observation. 

\noindent\textbf{\textit{Setup:\ \ }}
The system begins with 1000 warm-up tasks, followed by 200 user requests. The reward function weights are set to \( w_1 = 0.1 \) and \( w_2 = 0.2 \), respectively. 

\subsection{Task Offloading in Edge Computing}
Task offloading in edge computing delegates computational tasks from resource-constrained edge devices to more capable edge servers, thereby mitigating local power consumption while preserving acceptable processing latency. It is particularly crucial for managing image inference tasks on battery-powered edge devices, such as smartphones, where the computational demands of deep learning models can rapidly deplete limited energy resources. In our experiments, we explore task offloading for a variety of image-based inference tasks.

\noindent\textbf{\textit{Action Space:\ \ }}The action space \(\mathcal{A} = \{0, 1, 2, 3, 4\}\) includes five discrete options representing different offloading levels. Action \(a = 0\) denotes full local execution on the client, while \(a = 4\) corresponds to complete offloading to the server. Intermediate actions \(a = 1\), \(2\), and \(3\) indicate offloading 25\%, 50\%, and 75\% of the images, respectively. 

\noindent\textbf{\textit{State Representation:\ \ }}The state space \(\mathcal{S} \in \mathbb{R}^{10}\) captures essential runtime information, including
\begin{itemize}
    \item A one-hot encoded vector indicating the current task type;
    \item Number of images in the current task;
    \item Average image size, represented as a 2D vector \([w, h]\);
    \item Remaining battery level of the client;
    \item GPU working frequency on the client device.
\end{itemize}

\noindent\textbf{\textit{Reward Function:\ \ }}
The delay  $T_{\text{delay}}$ is measured as the total processing time from the beginning to the end of a task. The remaining battery \(B_{\text{remain}}\) is calculated based on the input voltage and current over time. The reward \(r\) at each step is
\begin{equation}\label{eq:edge_reward}
    r = -\left( w_1 \cdot T_{\text{delay}} + w_2 \cdot (1 - B_{\text{remain}}) \right).
\end{equation}

\noindent\textbf{\textit{Partial Observability:\ \ }}
Several factors affecting delay and energy consumption are not directly observable in the state representation. In particular, the instantaneous wireless network condition, the computational load of the edge server, and the thermal state of the client device remain hidden. 

\noindent\textbf{\textit{Setup:\ \ }}
The offloading task is built on the YOLOv11n series models~\cite{YOLOV11}. We set \( w_1 = 0.5 \) and \( w_2 = 1 \). The client device is initialized with a random power level in the range [60,100]. Each task consists of several images. the number of which is sampled from the set \(\{4, 8, 16, 24, 32\}\) with probabilities  \(\mathbf{p}_{\text{size}} = \{0.2, 0.2, 0.3, 0.2, 0.1\}\). The task type is randomly selected from five YOLOv11-based operations: detection, segmentation, classification, oriented bounding box (OBB) estimation, and pose estimation, based on the probability distribution \(\mathbf{p}_{\text{type}} = \{0.25, 0.2, 0.2, 0.3, 0.05\}\). Each trajectory is terminated when the edge device has processed 200 tasks or can no longer operate due to battery depletion.

\subsection{Channel Selection in Wi-Fi Communication}
Adaptive channel selection is a widely adopted in Wi-Fi to enhance communication efficiency. Particularly, the transmitter dynamically selects a channel from a predefined candidates for packet transmission at a fixed rate. However, channel quality varies over time due to interference, fading, and mobility. To track these fluctuations, the receiver maintains a Channel State Information (CSI) table to log the most recent observations for each channel. The CSI is only updated when a channel is actively utilized. 

\noindent\textbf{\textit{Action Space:\ \ }}The action space consists of \(n = 8\) discrete choices, \(\mathcal{A} = \{1, 2, \dots, n\}\), each representing the selection of a specific wireless channel for data transmission.

\noindent\textbf{\textit{State Representation:\ \ }}The state space, \(\mathcal{S} \in \mathbb{R}^{53 \times |\mathcal{A}|}\), is defined as a 53-dimensional vector of CSI values corresponding to all channels. Since only the CSI of the currently used channel is updated, the agent hinges on historical CSI values for unselected channels, making the problem partially observable and temporally dependent.

\noindent\textbf{\textit{Reward Function:\ \ }} At each step, the transmitter selects a new channel and starts sending packets at a fixed rate. The receiver collects incoming packets and tracks their sequence numbers. Once 100 packets (based on sequence numbers) have been received, the packet loss rate is calculated and used to define the reward
\begin{equation}
    r = -P_{\text{miss}}.
\end{equation}

\noindent\textbf{\textit{Partial Observability:\ \ }}
The CSI table stores the latest observation for each channel, while CSI values are updated only when a channel is selected. Consequently, the current quality of unselected channels, external interference from nearby wireless devices, and the underlying channel fading dynamics remain hidden.

\noindent\textbf{\textit{Setup:\ \ }}
A total of eight channels are pre-defined, including five channels in the 5~GHz band (5.17~GHz, 5.20~GHz, 5.27~GHz, 5.32~GHz, and 5.72~GHz), and three in the 2.4~GHz band (2.412~GHz, 2.442~GHz, and 2.472~GHz). Each trajectory consists of 200 steps.

\section{Evaluation}
This section describes the experiment setup, including real-world data collection, and presents experimental results to demonstrate the effectiveness of proposed MEO framework.

\subsection{Experimental Setup}
\subsubsection{Dataset Preparation}
We describe the process of collecting offline datasets for training across three use cases:

\noindent\textbf{Resource Allocation:} We instantiate the server resource allocation process in ORAN-Cloud using the Alibaba Cluster dataset\footnote{\url{https://github.com/alibaba/clusterdata}}, which serves as the workload source. We then collect offline training data by executing Greedy and Round-Robin policies, generating 1 million samples from each policy. The Greedy policy selects the server with the fewest pending tasks, promoting better load balancing. 

\begin{figure}[t]
    \centering
    \begin{subfigure}[b]{0.23\textwidth}
        \centering
        \includegraphics[width=0.9\linewidth]{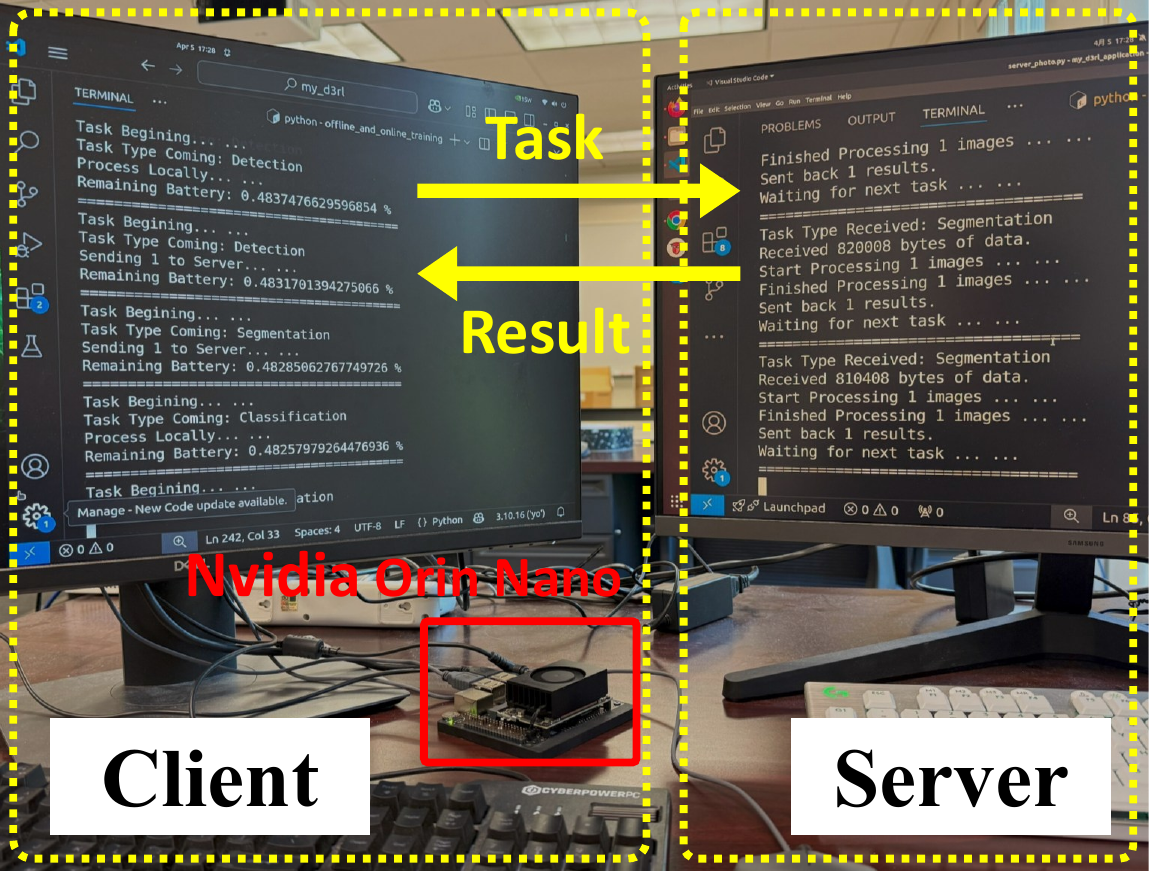}
        \label{fig:framework_edge}
    \end{subfigure}
    \begin{subfigure}[b]{0.23\textwidth}
        \centering
        \includegraphics[width=0.9\linewidth]{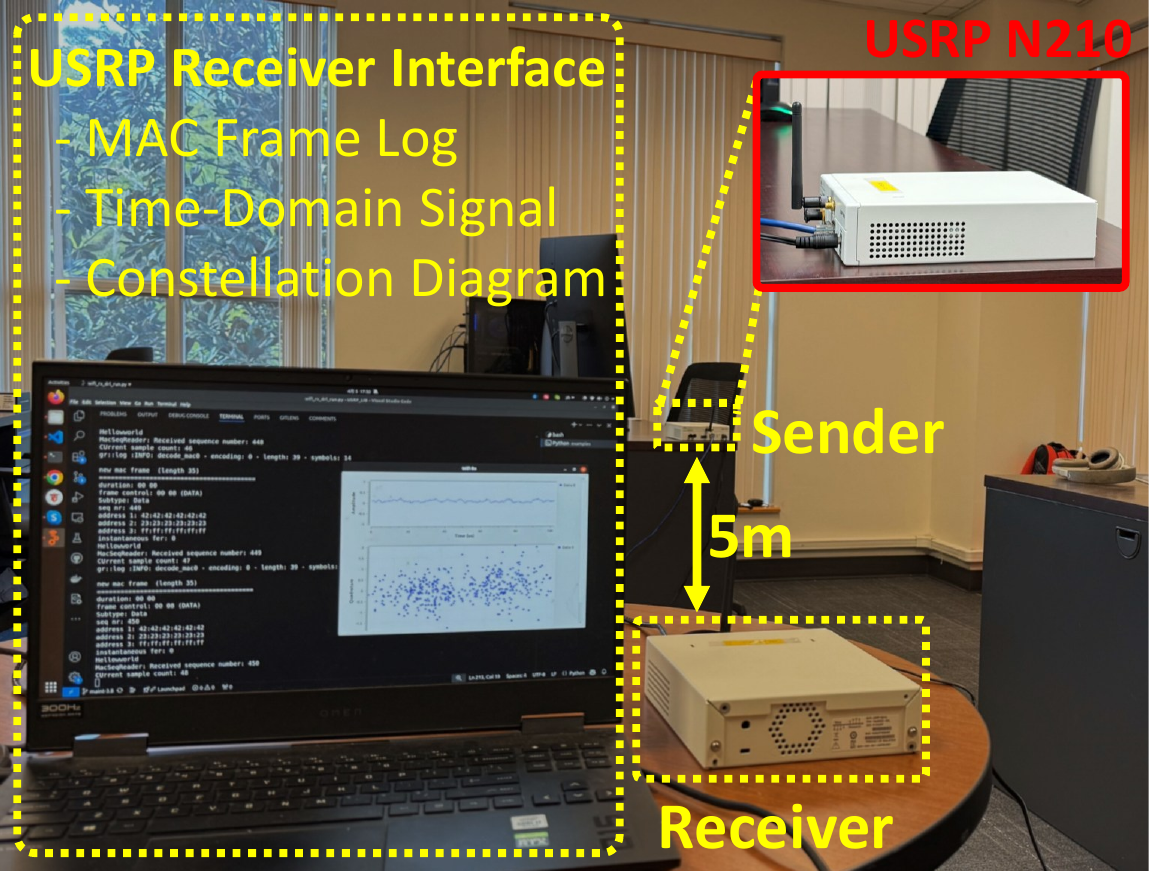}
        \label{fig:framework_usrp}
    \end{subfigure}
    \caption{Illustration of Task Offloading between edge client and server(\textit{left}) and Channel Selection for Wi-Fi signal transmission deployed on USRP (\textit{right}).}
    \Description{}
    \label{fig:applications}
\end{figure}

\noindent\textbf{Task Offloading:} We employ both Greedy and Round-Robin policies to collect 50{,}000 samples each. The Greedy strategy makes decisions based on the client's remaining battery level to extend device longevity. Data collection takes approximately 52 hours for the Greedy and 46 hours for Round-Robin. The overall task offloading between the edge client and server is illustrated in Fig.~\ref{fig:applications} (\textit{left}).

\noindent\textbf{Channel Selection:} We collect 153{,}400 samples using a Round-Robin policy over 7 days of continuous transmission. The channel selection process for Wi-Fi signal transmission deployed on USRP is illustrated in Fig.~\ref{fig:applications} (\textit{right}).

For K-means clustering, we set the cluster size to \(c=20{,}000\), \(c=1{,}000\), and \(c=2{,}000\) for Resource Allocation, Task Offloading, and Channel Selection, respectively. We then choose $k$ from the average cluster size so that pseudo-experts account for roughly 5\% of a cluster, which gives $k = 2$ in all three cases. For the different cluster sizes in Tab.~\ref{tab:offlineoptimal}, $k$ is adjusted accordingly so that the total number of selected samples stays constant.

\subsubsection{Algorithms Implementation}
For offline RL training, we focus mainly on CQL~\cite{kumar2020conservative}, implemented from the offline Deep RL library D3RL \footnote{https://github.com/takuseno/d3rlpy}. We also use two trajectory-based algorithms, CLUE~\cite{liu2023clue} and OTR~\cite{luo2023optimaltransportofflineimitation}, as comparison baselines. These baselines serve as method-level ablations of the reward stage: CQL applies no relabeling, OTR relabels without a CVAE representation, and CLUE relabels from a CVAE representation built on trajectory-level expert selection, where the expert trajectories are chosen by top returns and kept at the same size as our expert set for fairness.

\subsubsection{Enhanced CVAE}
The action embedding dimension is set to 8, corresponding to the discrete action space after embedding layer. The hidden dimension is configured to 128 for Resource Allocation and 64 for others. We set \(\lambda_{reg}\) to 40.0 to scale the regularization loss (Eq. \ref{Eq:regLoss}) and \(\lambda_{neg}\) to 5.0 to scale the negative pair loss (Eq. \ref{Eq:L_neg}). The \(m_e\) is specified by the margin parameter and is set to 0.3 (Eq. \ref{Eq:L_neg}). We perform the training of the model over 20{,}000 iterations, with a batch size of 64. The learning rate (\(lr\)) is configured at 3e-4 with a weight decay of 0.0001 to regularize the weights during training.

\subsubsection{Offline Training}
We configure the offline training with following hyperparameters. The learning rate is 1e-4 with a batch size of 64. In CQL, the conservative weight \(\alpha\)  is  1.0. The discount factor \(\gamma\) is  0.9 for task offloading and 0.8 for others. The Adam optimizer employs a weight decay of 5e-5. The neural network comprises a two-layer encoder with hidden dimensions of 64, incorporating batch normalization and a dropout rate of 0.2. The Q-function adopts a quantile regression approach with 100 quantiles. Training is performed using five distinct random seeds, with five trials per seed, resulting in 25 trials. The reported results are averaged across all trials to ensure robustness and reliability.

\subsection{Results on Resource Allocation}
We evaluate MEO for resource allocation in ORAN-Cloud. For the offline training phase, we undertake a substantial $5e5$ steps to comprehensively train the models. Subsequently, the models are fine-tuned for an additional $150{,}000$ steps in the online system to further optimize their performance.

\subsubsection{Offline Performance and Adaptability of MEO}
Tab.~\ref{tab:offline_adaptive} reports the offline performance of MEO on ORAN-Cloud. MEO improves the average return on both datasets with CQL~\cite{kumar2020conservative}, as well as with DDQN~\cite{van2016deep}, BCQ~\cite{fujimoto2019off}, and SAC~\cite{haarnoja2019soft}~\footnote{These algorithms are selected based on the D3RL library's recommendation for discrete action spaces. Other offline algorithms, such as IQL~\citep{kostrikov2021implicit}, are primarily designed for continuous action spaces and are therefore not applicable to the current setting.}, with the gain significant in every case. These results demonstrate that MEO can be effectively integrated with different RL algorithms.
\begin{table}[t]
    \caption{Offline performance of MEO across RL backbones.
    The $\uparrow$ indicates improvement over the base method, and $p$-values are
    calculated using the independent $t$-test.}
    \centering
    \begin{tabular}{lcccc}
    \toprule
                              & DDQN   & BCQ    & SAC    & CQL \\
    \midrule
    Greedy                    & -72.48 & -71.85 & -72.04 & -71.38 \\
    \textbf{\quad +MEO}       & \textbf{-71.12$\uparrow$} & \textbf{-70.34$\uparrow$} & \textbf{-70.46$\uparrow$} & \textbf{-70.73$\uparrow$} \\
    $p$-value                 & 5.96e-3 & 2.56e-3 & 2.07e-5 & 4.17e-2 \\
    \midrule
    Round-Robin               & -70.88 & -72.68 & -72.40 & -72.44 \\
    \textbf{\quad +MEO}       & \textbf{-68.48$\uparrow$} & \textbf{-70.62$\uparrow$} & \textbf{-66.96$\uparrow$} & \textbf{-69.48$\uparrow$} \\
    $p$-value                 & 2.32e-12 & 6.01e-9 & 1.12e-39 & 8.33e-15 \\
    \bottomrule
    \end{tabular}
    \label{tab:offline_adaptive}
\end{table}

\subsubsection{Parameters Analysis}
\begin{table}[t]
\caption{Effect of the clustering size $c$ and the scaling factor $\tau$ on CQL+MEO, on the Round-Robin dataset.}
\centering
\begin{tabular}{lcccc}
\toprule
& \textbf{$c=2{,}500$} & \textbf{$c=5{,}000$} & \textbf{$c=10{,}000$} & \textbf{$c=20{,}000$} \\
\midrule
\textbf{$\tau$} = 0.1 & -71.907 & -71.305 & -70.843 & -71.127 \\
\textbf{$\tau$} = 0.3 & -70.664 & -69.986 & -69.841 & \textbf{-69.482} \\
\textbf{$\tau$} = 0.5 & -70.972 & -69.991 & -69.980 & -69.744 \\
\bottomrule
\end{tabular}
\label{tab:offlineoptimal}
\end{table}

Tab.~\ref{tab:offlineoptimal} presents a detailed comparison of performance under different clustering sizes $c$ and scaling factor $\tau$. As the number of clusters increases, the integration of MEO consistently leads to improved returns. The largest improvement is observed when $\tau = 0.3$ with the cluster size $c = 20{,}000$. These results indicate that MEO can effectively exploit the clustering structure of the dataset to improve policy performance.
\begin{figure}[t]
    \centering
    \begin{subfigure}[b]{0.235\textwidth}
        \centering
        \includegraphics[width=0.99\linewidth]{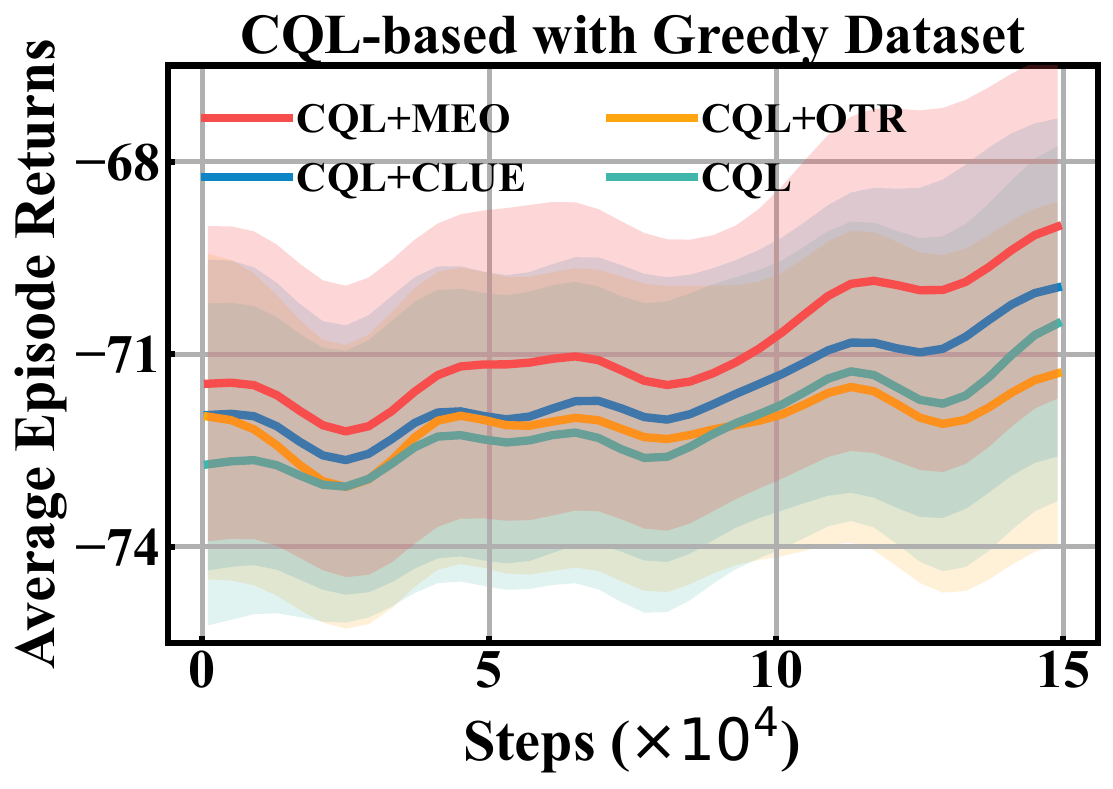}
        \label{fig:online_CQL_greedy}
    \end{subfigure}
    \begin{subfigure}[b]{0.235\textwidth}
        \centering
        \includegraphics[width=0.99\linewidth]{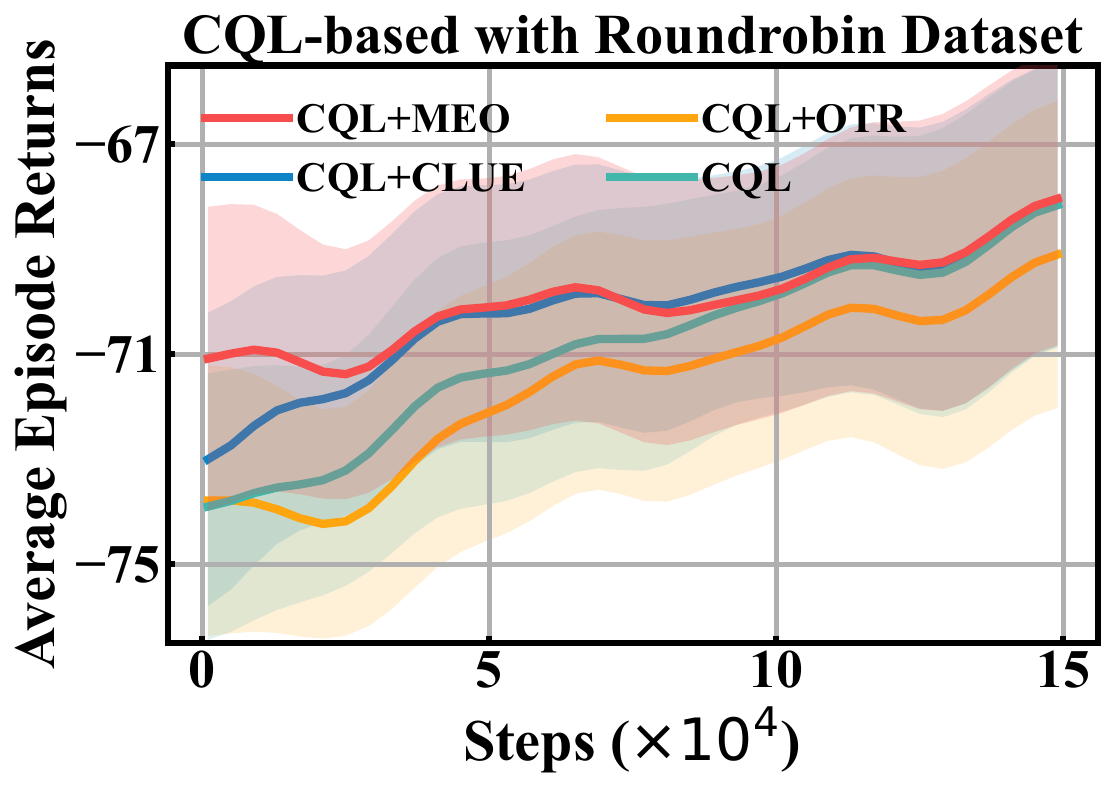}
        \label{fig:online_CQL_roundrobin}
    \end{subfigure}
    \caption{Fine-tuning performances in Resource Allocation. Shaded regions denote one standard error over 5 seeds.}
    \Description{}
    \label{fig:finetune}
\end{figure}
\subsubsection{Online Fine-Tuning}
\begin{figure}[t]
    \centering
    \includegraphics[width=0.43\textwidth]{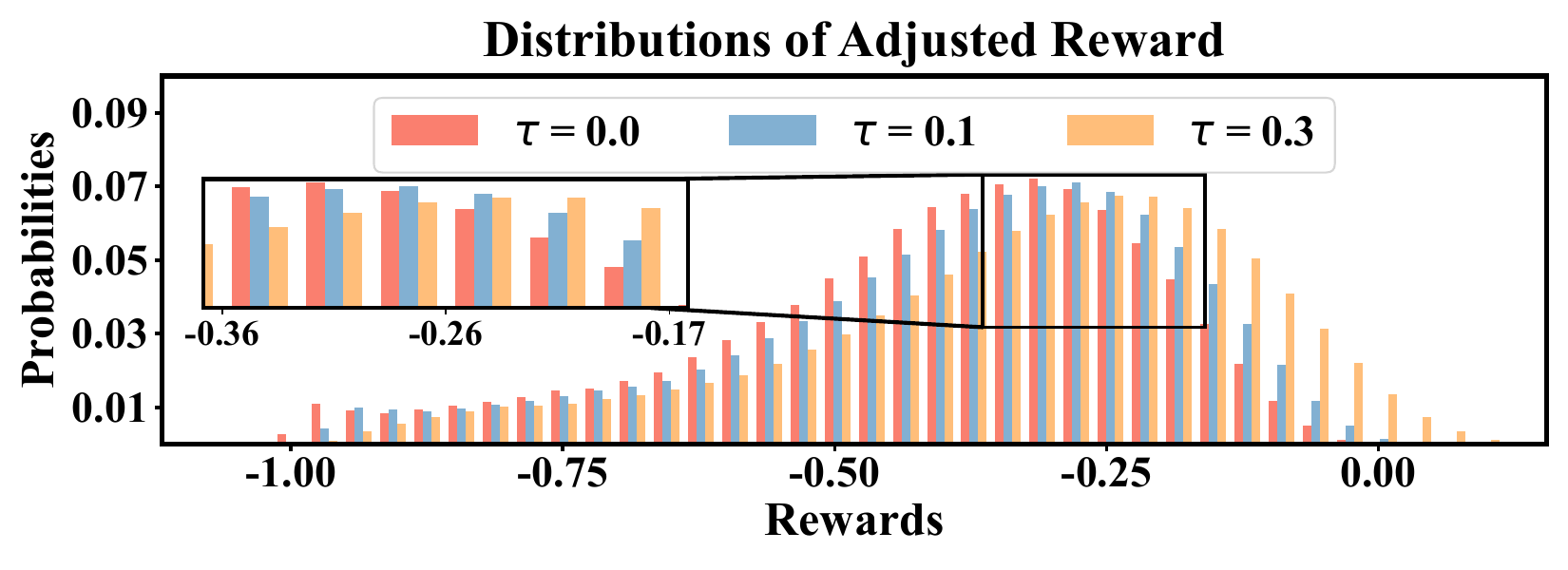}
    \caption{Adjusted reward distributions for different \(\tau\) scales in the Greedy dataset.}
    \Description{}
    \label{fig:distribution}
\end{figure}

Beyond the offline evaluation above, we additionally examine how an offline-trained policy adapts when limited online interaction becomes available. Fig.~\ref{fig:finetune} compares the performance of MEO against baselines during online fine-tuning. Here, $t = 0$ corresponds to the policies obtained after offline training, so the methods start from different levels, and MEO begins highest on both datasets. Additionally, Fig.~\ref{fig:distribution} shows how $\tau$ (see Eq.~\ref{Eq:tau}) reshapes the reward distribution. As $\tau$ increases from $0.0$ to $0.3$, the peak shifts from around $-0.3$ to $-0.2$, confirming that the compensable reward produces a more informative learning signal.

\begin{figure}[H]
    \centering
    \begin{subfigure}[b]{0.235\textwidth}
        \centering
        \includegraphics[width=0.98\linewidth]{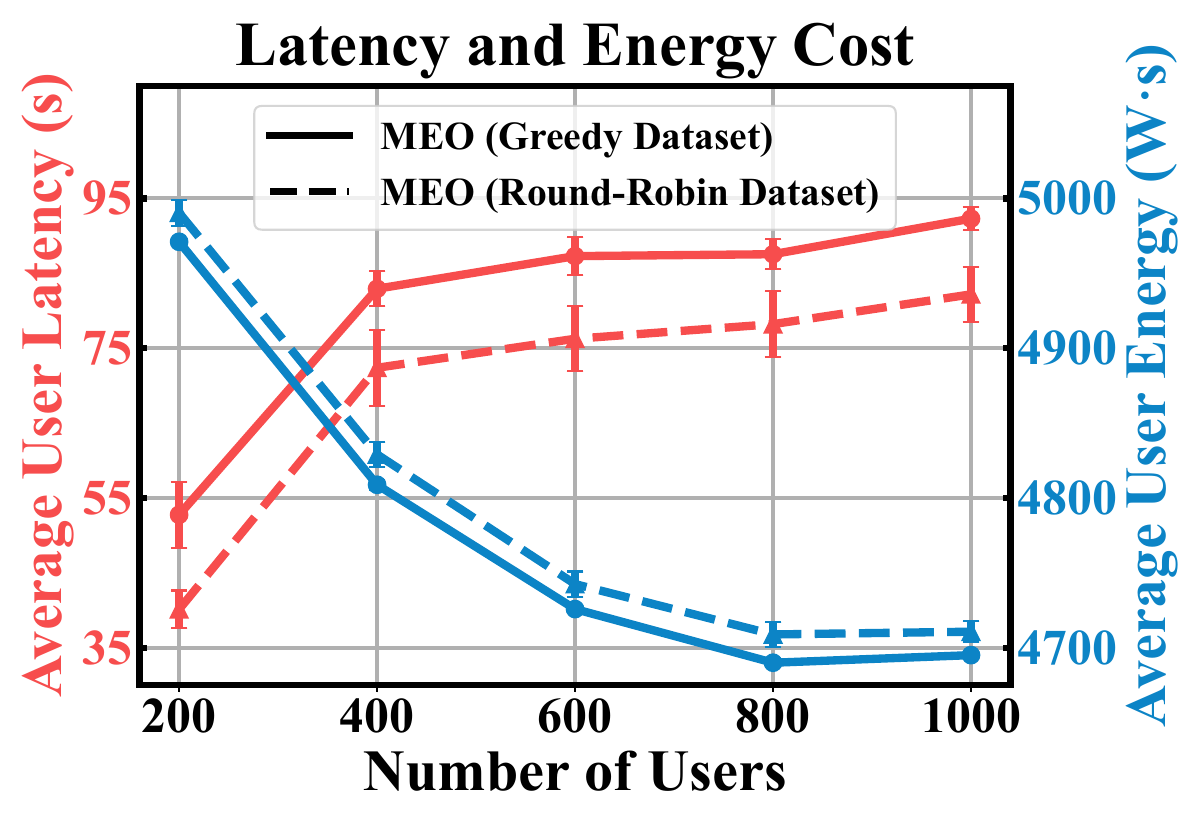}
        \label{fig:power}
    \end{subfigure}
    \begin{subfigure}[b]{0.235\textwidth}
        \centering
        \includegraphics[width=0.98\linewidth]{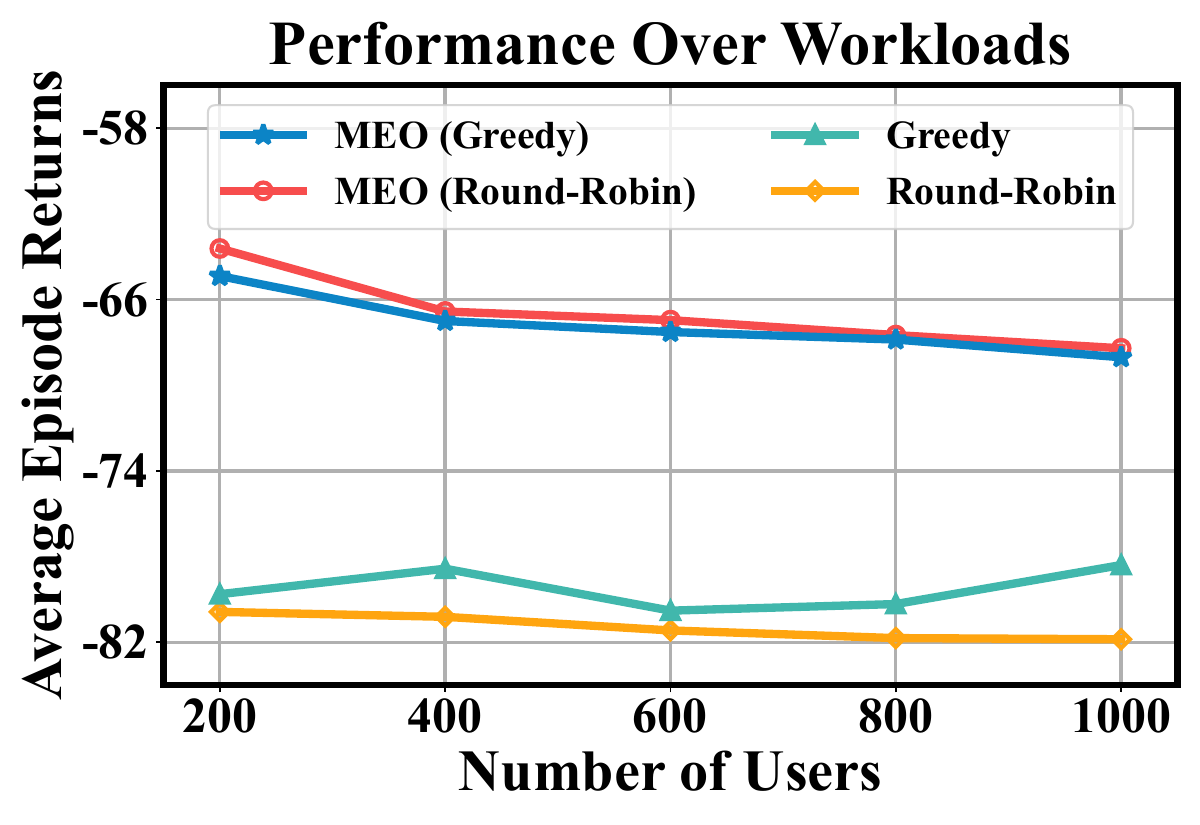}
        \label{fig:workload}
    \end{subfigure}
    \caption{Latency and energy profiles (\textit{left}), and behavior policy across workloads (\textit{right}).}    
    \Description{}
    \label{fig:workloads}
\end{figure}

Beyond numerical performance, we further examine how MEO affects system-level behavior under different workloads. Fig.~\ref{fig:workloads} \textit{(left)} shows that the two training datasets lead to distinct latency and energy profiles as the number of users grows, while \textit{(right)} shows that MEO improves over the underlying behavior policy.

\subsubsection{Ablation for CVAE model}
\begin{figure}[H]
    \centering
    \begin{subfigure}[b]{0.238\textwidth}
        \centering
        \includegraphics[width=0.98\linewidth]{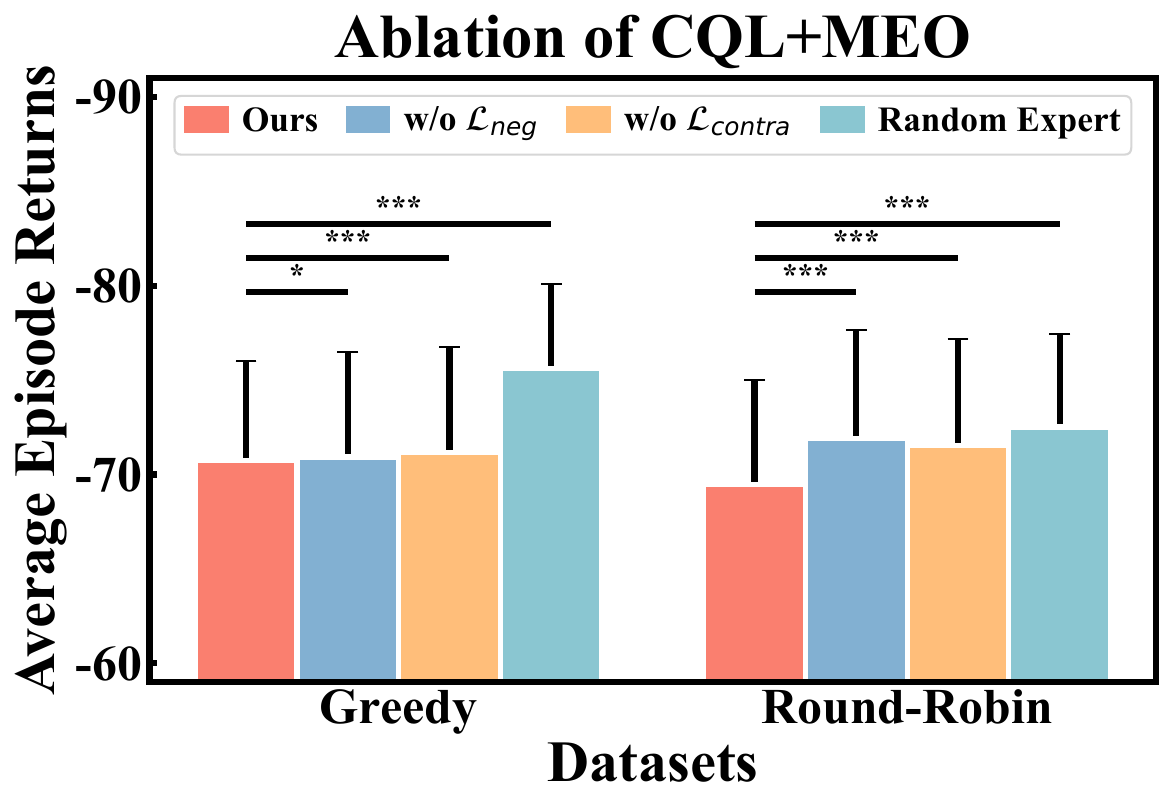}
        \label{fig:ablation-cql}
    \end{subfigure}
    \begin{subfigure}[b]{0.2292\textwidth}
        \centering
        \includegraphics[width=0.98\linewidth]{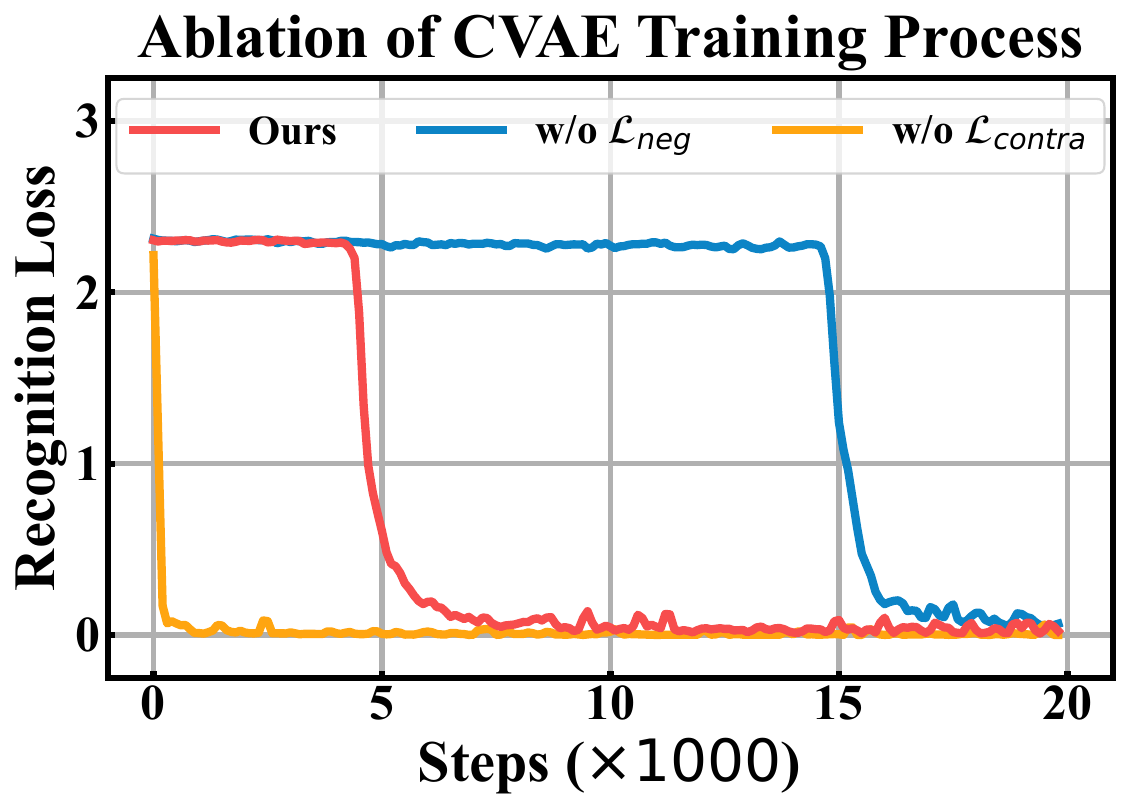}
        \label{fig:ablation-cvae}
    \end{subfigure}
    \caption{Ablation study across datasets (\textit{left}), along with CVAE training convergence (\textit{right}) on Round-Robin dataset. (\(^*p\text{-value}<0.05\), \(^{**}p\text{-value}<0.01\), \(^{***}p\text{-value}<0.001\))}
    \Description{}
    \label{fig:ablation}
\end{figure}

The ablation isolates the three components of MEO. As shown in Fig.~\ref{fig:ablation}(\textit{left}), removing $\mathcal{L}_{\text{neg}}$ or $\mathcal{L}_{\text{contra}}$ both degrade performance, and replacing cluster-wise Top-$K$ with random expert selection degrades it the most. Moreover, Fig.~\ref{fig:ablation}(\textit{right}) illustrates the convergence behavior of the training.

\subsection{Results on Task Offloading}
We evaluate MEO in a practical image task offloading, as illustrated in Fig.~\ref{fig:applications} (\textit{left}). We perform a total of $2e4$ offline training steps given the limited dataset size of $5e4$ samples. Notably, this dataset is significantly smaller than typical offline datasets.

\subsubsection{Comparison with Baselines}
Tab.~\ref{tab:offlineresults_edge} presents the average trajectory returns of various methods trained on Greedy and Round-Robin datasets. From Tab.~\ref{tab:offlineresults_edge}, MEO consistently outperforms all baselines across both datasets, which indicates the effectiveness and stability of our method. In addition, our results on the Round-Robin dataset are generally better than those on the Greedy dataset across all methods. This is likely because Round-Robin generates more uniformly distributed actions and experiences, which offers a better initial coverage for offline learning. It is worth noting that both CLUE and OTR hinge on identifying expert trajectories, which limits their performance when expert samples are sparse or unreliable. In contrast, MEO is more robust and better suited to limited datasets via leveraging state-based advantage estimation.  
\begin{table}[H]
    \caption{Average trajectory returns comparison among baselines in Greedy and Round-Robin datasets.}
    \centering
    \begin{tabular}{lcc}
    \toprule
     & Greedy & Round-Robin \\  
    \midrule
     CQL          & -187.8068 \footnotesize{\( \pm \) 1.3229} & -127.2304 \footnotesize{\( \pm \) 1.9439} \\
     CQL+OTR    & -192.0710 \footnotesize{\( \pm \) 2.6662} & -128.0848 \footnotesize{\( \pm \) 2.4420} \\ 
     CQL+CLUE   & -215.0105 \footnotesize{\( \pm \) 18.3347} & -129.3592 \footnotesize{\( \pm \) 3.1457} \\
     \textbf{CQL+MEO}   & \textbf{-181.9689} \footnotesize{\( \pm \) 1.0237} & \textbf{-125.1219} \footnotesize{\( \pm \) 1.6114} \\ 
    \bottomrule
    \end{tabular}
    \label{tab:offlineresults_edge}
\end{table}

\subsubsection{Parameter Sensitivity}
\begin{figure}[t]
  \centering
  \begin{subfigure}[t]{0.24\linewidth}
    \centering
    \includegraphics[width=\linewidth]{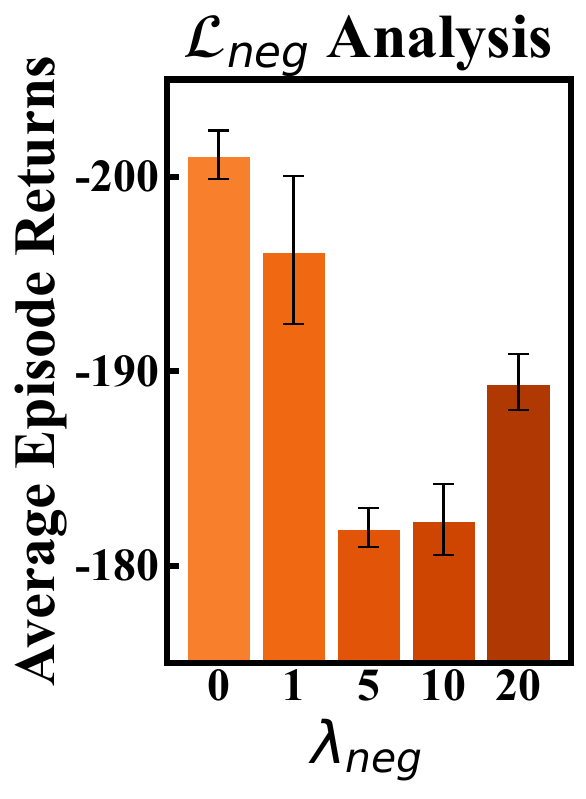}
  \end{subfigure}%
  \hspace{0.06in}
  \begin{subfigure}[t]{0.24\linewidth}
    \centering
    \includegraphics[width=\linewidth]{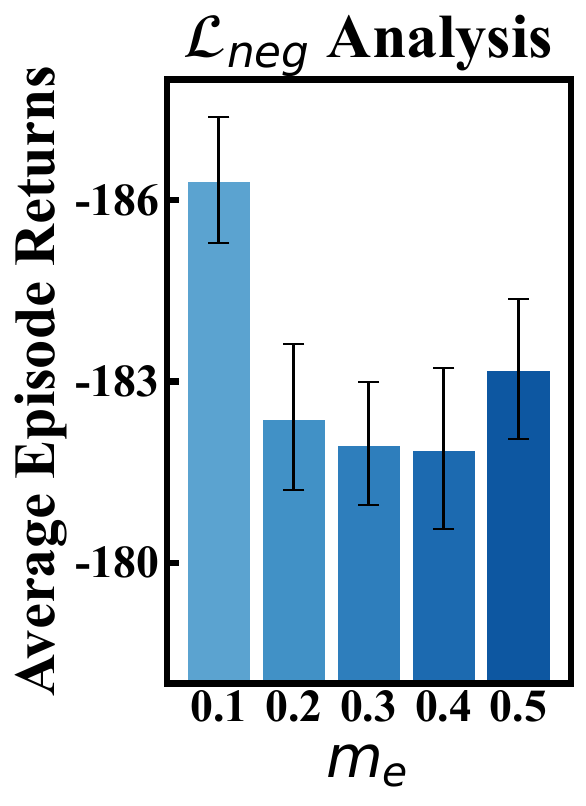}
  \end{subfigure}
  \hspace{0.06in}
  \begin{subfigure}[t]{0.45\linewidth}
    \centering
    \includegraphics[width=\linewidth]{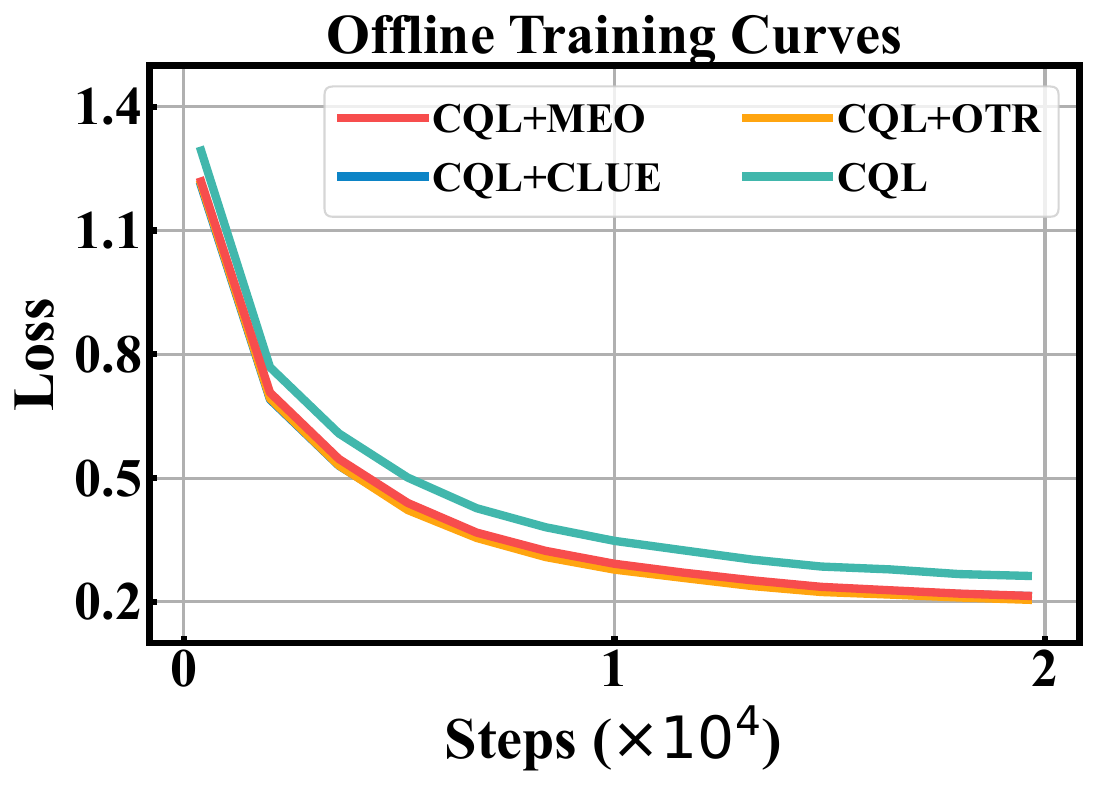}
  \end{subfigure}%
  \caption{Parameter sensitivity analysis of $\mathcal{L}_{\text{neg}}$ under different $\lambda_{\text{neg}}$ (\textit{left}) and $m_e$ (\textit{middle}), along with the offline training curves (\textit{right}) on the Greedy dataset.}
  \label{fig:edge_ablation}
\end{figure}

The parameter analysis illustrates the effect of different $\lambda_{neg}$ and $m_e$ in $\mathcal{L}_{\text{neg}}$ (Eq.~\ref{Eq:L_neg}). As shown in Fig.~\ref{fig:edge_ablation}(\textit{left}) and (\textit{middle}), the performance results suggest a non-trivial trade-off: Overly small values provide limited guidance, while large values can degrade performance due to over-regularization or distorted reward signals. Nevertheless, Fig.~\ref{fig:edge_ablation}(\textit{right}) presents the training curves of different methods, showing that the proposed method exhibits stable convergence during offline training.

\begin{figure*}[htbp]
  \centering
  \begin{minipage}[t]{0.32\linewidth}
    \centering
    \includegraphics[width=1\textwidth]{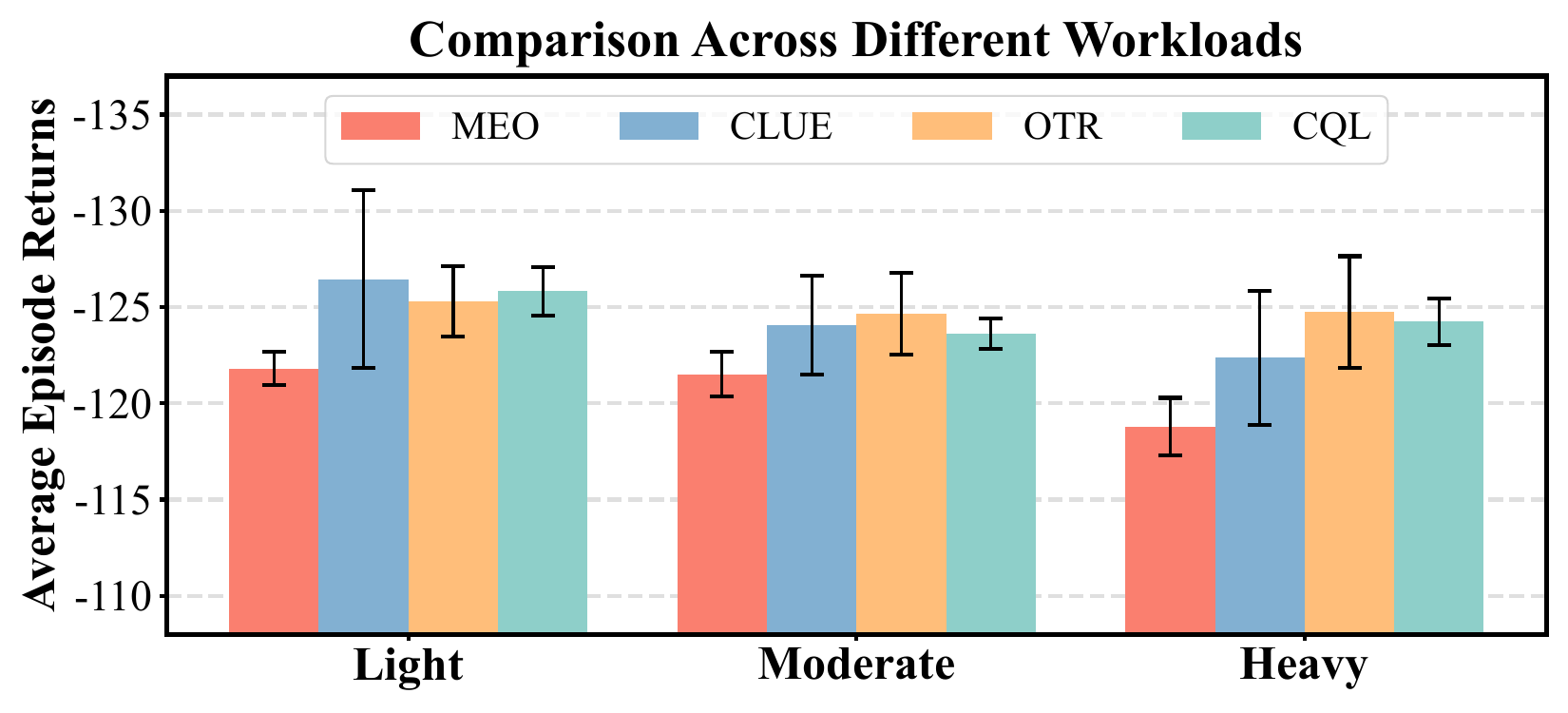}
    \caption{Performance of Task Offloading under three unseen distributions.}
    \label{fig:workloads_edge}
  \end{minipage}%
  \hspace{0.06in}
  \begin{minipage}[t]{0.32\linewidth}
    \centering
    \includegraphics[width=1\textwidth]{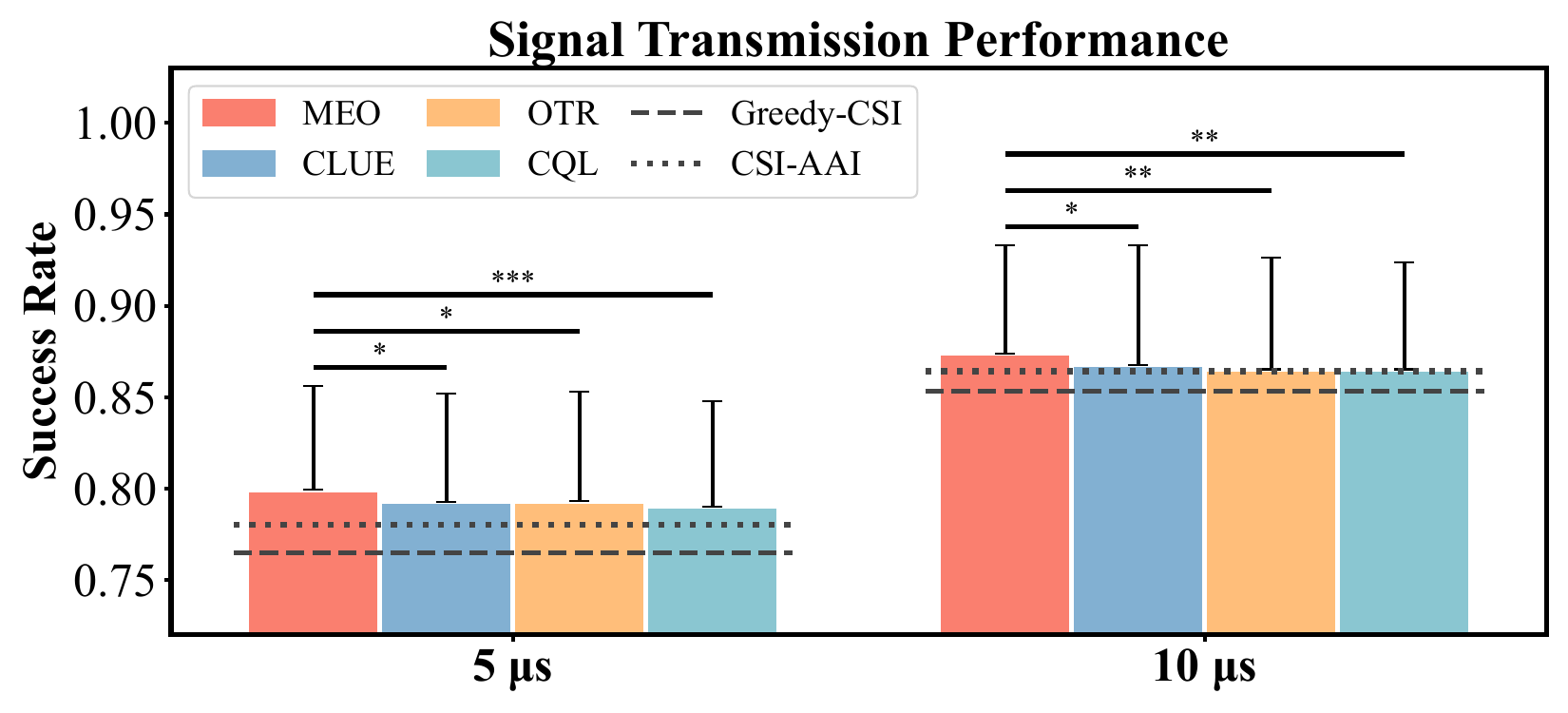}
    \caption{Performance of Channel Selection on $5\,\mu s$ and $10\,\mu s$ intervals.}
    \label{fig:results_usrp}
  \end{minipage}%
    \hspace{0.06in}
  \begin{minipage}[t]{0.32\linewidth}
    \centering
    \includegraphics[width=1\textwidth]{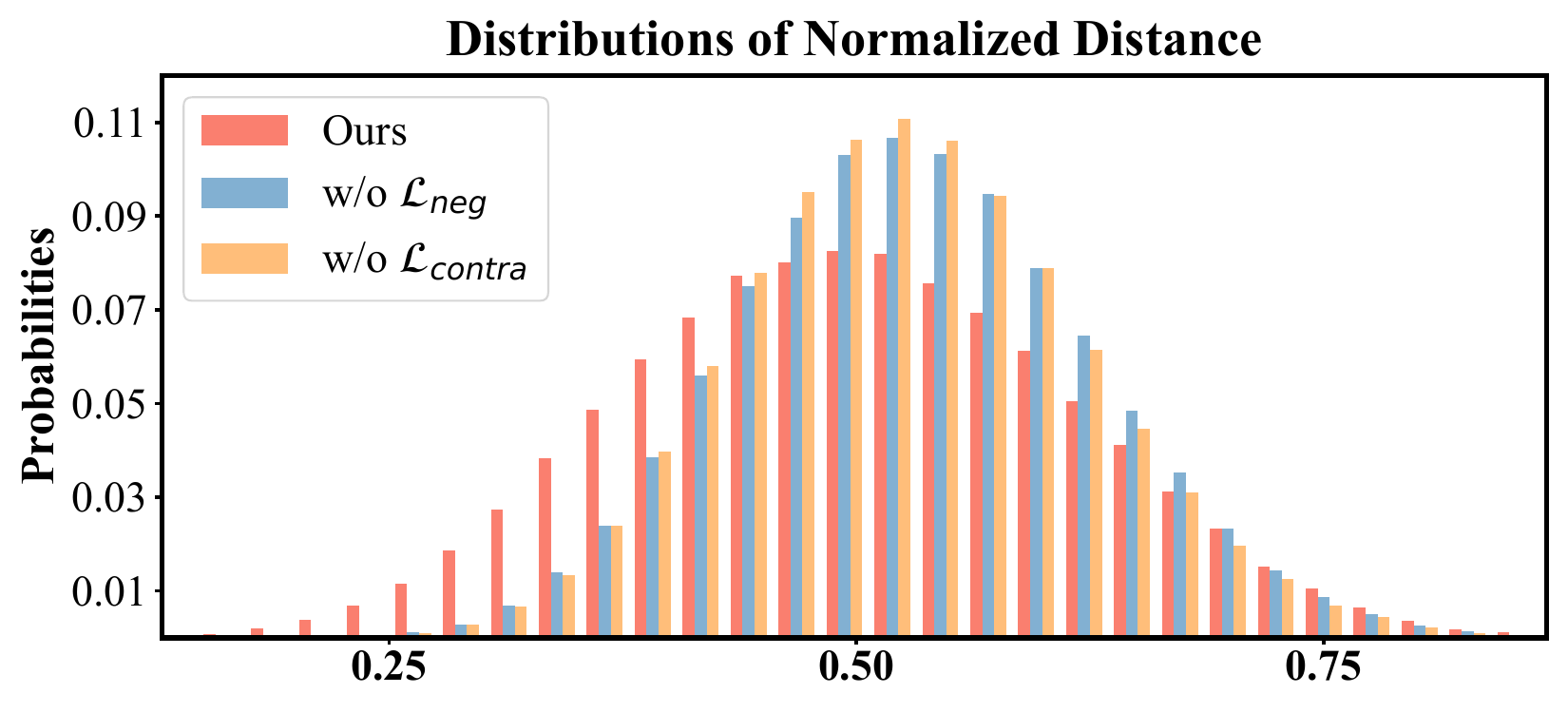}
    \caption{Illustration of normalized distance distributions.}
    \Description{}
    \label{fig:usrp_distance}
  \end{minipage}
\end{figure*}

\subsubsection{Performance under Shifted Workload Distributions}
To evaluate the generalization ability of MEO and baselines beyond the training data distribution, three OOD task workloads characterized by different image size sampling probabilities are examined. Specifically, the sampling probabilities are defined as follows: \(\mathbf{p}_{\text{size}} = \{0.3, 0.3, 0.2, 0.1, 0.1\}\) (expected size: 12.4) for the \textit{light workload}; \(\mathbf{p}_{\text{size}} = \{0.15, 0.3, 0.15, 0.3, 0.1\}\) (expected size: 15.8) for the \textit{moderate workload};  \(\mathbf{p}_{\text{size}} = \{0.2, 0.1, 0.2, 0.1, 0.4\}\) (expected size: 20.0) for the \textit{heavy workload}. These distributions deviate significantly from the one used during offline dataset collection, thereby introducing substantial shifts in workload characteristics. As shown in Fig.~\ref{fig:workloads_edge}, MEO consistently surpasses all baselines across these unseen workload settings with  the highest average returns in all three cases. The result highlights MEO’s ability to generalize effectively to dynamic and heterogeneous task distributions in real-world task offloading. 

\subsection{Results on Channel Selection}
We deploy MEO for channel selection in  Wi-Fi transmission on USRP as shown in Fig.~\ref{fig:applications} (\textit{right}). The receiver decodes the signal in real time. The signal quality is monitored via the MAC frame logs, time-domain signals, and the constellation diagram. We perform a total of $5e4$ offline training steps given the dataset size of $153{,}400$.

\subsubsection{Comparison with Baselines}
The packet success rate is adopted as the evaluation metric, measured via real-time packet reception over 1{,}000 environment steps per seed across 5 random seeds, with significance assessed by a paired $t$-test over the seeds. As shown in Fig.~12, MEO attains the highest success rate at both transmission intervals, 0.7992 at $5\,\mu s$ and 0.8739 at $10\,\mu s$, and the margin over all baselines is significant in every case. We also compare against two non-RL heuristics: Greedy-CSI, which takes the channel with the highest stored value, and a CSI-aging-aware index (CSI-AAI), which adds an age-based uncertainty bonus for outdated CSI. Only MEO exceeds both heuristics at either interval, which is consistent with our earlier observation that expert trajectories are unreliable when returns are shaped by unobserved channel dynamics.

\subsubsection{Ablation Study on Distance Distribution in CVAE's Latent Space}
To explore the effectiveness of compensable reward mechanism, we further analyze the latent distance distributions. Fig.~\ref{fig:usrp_distance} visualizes the normalized distances between expert and non-expert samples in the latent space. Obviously, MEO yields a broader distance distribution compared to the ablated baselines, which implies a more expressive measurement of behavioral discrepancy. 

\section{Related Works}
\textbf{Offline reinforcement learning.} Offline RL aims to learn decision policies from previously collected datasets without additional interaction with the environment~\cite{zeng2023demonstrations, li2024accelerating}. To improve learning stability from static datasets, existing studies have explored a variety of approaches, including conservative policy learning, imitation learning, and trajectory generation~\cite{ke2021grasping, foster2024behavior, wang2026trajectory}. A related line of work leverages expert samples or high-return trajectories to guide policy learning through optimal-transport alignment~\cite{luo2023optimaltransportofflineimitation, fu2024robot} or reward distillation~\cite{chaudhary2025from}. Several studies further employ representation learning to capture behavioral patterns in the dataset, enabling the model to distinguish informative behaviors from suboptimal ones~\cite{liu2023clue, zhao2025beyondexpert}. However, all of these approaches require a reliable expert reference at the outset, and address how to propagate it rather than how to obtain it. In practice, real-world datasets are collected from complex systems and often contain mixed-quality behaviors without explicit expert annotations, making it challenging to identify reliable expert signals for effective policy learning.

\textbf{Reinforcement learning in wireless networks.}
RL has been increasingly explored for wireless network optimization tasks, like in radio resource management~\citep{zangooei2023reinforcement}, and network control~\cite{zhao2024mobility}, though most studies adopt online frameworks requiring continuous interaction during training~\cite{fiandrino2023explora}. Offline RL remains comparatively underexplored, with recent work adapting existing algorithms or offline pipelines to wireless scenarios~\cite{bonati2023openran, yang2024advancing}. However, wireless environments are inherently partially observable because network conditions are only partially observed through limited measurements~\cite{liu2023optimizing, wang2025beyond}. Existing remedies act at the policy level, conditioning on observation histories through recurrent architectures~\cite{pmlr-v162-ni22a} or inferring an explicit belief state~\cite{han2020variational}. Offline learning admits an earlier intervention, since hidden factors distort the training data before any policy is fit. Our work therefore acts at the data level, mining reliable expert signals from imperfect wireless datasets and leveraging them to guide offline policy learning under partial observability, which is complementary to and can be combined with these policy-level designs.

\section{Conclusions}
In this paper, we study offline reinforcement learning for wireless networks, where imperfect datasets and partial observability make reliable policy learning challenging. To address this challenge, we propose MEO, an offline RL framework that mines reliable expert signals from wireless datasets through cluster-wise expert identification, contrastive representation learning, and compensable reward construction. Experiments across multiple wireless network tasks show that MEO consistently outperforms existing baselines. Moreover, in the task offloading scenario with only 50,000 samples, MEO still achieves substantial improvements over all baselines. These results highlight the practicality of MEO for real-world wireless systems with limited expert data, which provides a promising direction for reliable offline RL in wireless network management.

\newpage

\bibliographystyle{ACM-Reference-Format}
\bibliography{sample-base}

@article{haarnoja2019soft,
  title={Soft actor-critic algorithms and applications},
  author={Haarnoja, Tuomas and Zhou, Aurick and Hartikainen, Kristian and Tucker, George and Ha, Sehoon and Tan, Jie and Kumar, Vikash and Zhu, Henry and Gupta, Abhishek and Abbeel, Pieter and others},
  journal={arXiv:1812.05905},
  year={2019}
}

@article{kumar2020conservative,
  title={Conservative q-learning for offline reinforcement learning},
  author={Kumar, Aviral and Zhou, Aurick and Tucker, George and Levine, Sergey},
  journal={Advances in Neural Information Processing Systems},
  volume={33},
  pages={1179--1191},
  year={2020}
}

@inproceedings{van2016deep,
  title={Deep reinforcement learning with double q-learning},
  author={Van Hasselt, Hado and Guez, Arthur and Silver, David},
  booktitle={Proceedings of the AAAI conference on artificial intelligence},
  volume={30},
  number={1},
  year={2016}
}

@article{bonati2023openran,
  title={OpenRAN Gym: AI/ML development, data collection, and testing for O-RAN on PAWR platforms},
  author={Bonati, Leonardo and Polese, Michele and D’Oro, Salvatore and Basagni, Stefano and Melodia, Tommaso},
  journal={Computer Networks},
  volume={220},
  pages={109502},
  year={2023},
  publisher={Elsevier}
}

@article{fiandrino2023explora,
  title={EXPLORA: AI/ML EXPLainability for the Open RAN},
  author={Fiandrino, Claudio and Bonati, Leonardo and D'Oro, Salvatore and Polese, Michele and Melodia, Tommaso and Widmer, Joerg},
  journal={Proceedings of the ACM on Networking},
  volume={1},
  number={CoNEXT3},
  pages={1--26},
  year={2023},
  publisher={ACM New York, NY, USA}
}

@inproceedings{liu2023clue,
  title={Clue: Calibrated latent guidance for offline reinforcement learning},
  author={Liu, Jinxin and Zu, Lipeng and He, Li and Wang, Donglin},
  booktitle={Conference on Robot Learning},
  pages={906--927},
  year={2023},
  organization={PMLR}
}

@article{sohn2015learning,
  title={Learning structured output representation using deep conditional generative models},
  author={Sohn, Kihyuk and Lee, Honglak and Yan, Xinchen},
  journal={Advances in neural information processing systems},
  volume={28},
  year={2015}
}

@inproceedings{liu2017hierarchical,
  title={A hierarchical framework of cloud resource allocation and power management using deep reinforcement learning},
  author={Liu, Ning and Li, Zhe and Xu, Jielong and Xu, Zhiyuan and Lin, Sheng and Qiu, Qinru and Tang, Jian and Wang, Yanzhi},
  booktitle={2017 IEEE 37th international conference on distributed computing systems (ICDCS)},
  pages={372--382},
  year={2017},
  organization={IEEE}
}

@inproceedings{bonati2022openrangym,
  author = {Bonati, Leonardo and Polese, Michele and D'Oro, Salvatore and Basagni, Stefano and Melodia, Tommaso},
  title = {OpenRAN Gym: An Open Toolbox for Data Collection and Experimentation with AI in O-RAN},
  booktitle = {Proc. of IEEE WCNC Workshop on Open RAN Architecture for 5G Evolution and 6G},
  address={Austin, TX, USA},
  month={April},
  year = {2022}
}

@article{yang2024offline,
  title={Offline reinforcement learning for wireless network optimization with mixture datasets},
  author={Yang, Kun and Shi, Chengshuai and Shen, Cong and Yang, Jing and Yeh, Shu-Ping and Sydir, Jaroslaw J},
  journal={IEEE Transactions on Wireless Communications},
  volume={23},
  number={10},
  pages={12703--12716},
  year={2024},
  publisher={IEEE}
}

@article{levine2020offline,
  title={Offline reinforcement learning: Tutorial, review, and perspectives on open problems},
  author={Levine, Sergey and Kumar, Aviral and Tucker, George and Fu, Justin},
  journal={arXiv preprint arXiv:2005.01643},
  year={2020}
}

@article{yan2024simple,
  title={A simple solution for offline imitation from observations and examples with possibly incomplete trajectories},
  author={Yan, Kai and Schwing, Alex and Wang, Yu-Xiong},
  journal={Advances in Neural Information Processing Systems},
  volume={36},
  year={2024}
}

@misc{luo2023optimaltransportofflineimitation,
      title={Optimal Transport for Offline Imitation Learning}, 
      author={Yicheng Luo and Zhengyao Jiang and Samuel Cohen and Edward Grefenstette and Marc Peter Deisenroth},
      year={2023},
      eprint={2303.13971},
      archivePrefix={arXiv},
      primaryClass={cs.LG},
      url={https://arxiv.org/abs/2303.13971}, 
}

@article{li2024accelerating,
  title={Accelerating exploration with unlabeled prior data},
  author={Li, Qiyang and Zhang, Jason and Ghosh, Dibya and Zhang, Amy and Levine, Sergey},
  journal={Advances in Neural Information Processing Systems},
  volume={36},
  year={2024}
}

@inproceedings{ke2021grasping,
  title={Grasping with chopsticks: Combating covariate shift in model-free imitation learning for fine manipulation},
  author={Ke, Liyiming and Wang, Jingqiang and Bhattacharjee, Tapomayukh and Boots, Byron and Srinivasa, Siddhartha},
  booktitle={2021 IEEE International Conference on Robotics and Automation (ICRA)},
  pages={6185--6191},
  year={2021},
  organization={IEEE}
}

@article{hippalgaonkar2023knowledge,
  title={Knowledge-integrated machine learning for materials: lessons from gameplaying and robotics},
  author={Hippalgaonkar, Kedar and Li, Qianxiao and Wang, Xiaonan and Fisher III, John W and Kirkpatrick, James and Buonassisi, Tonio},
  journal={Nature Reviews Materials},
  volume={8},
  number={4},
  pages={241--260},
  year={2023},
  publisher={Nature Publishing Group UK London}
}

@inproceedings{fujimoto2019off,
  title={Off-policy deep reinforcement learning without exploration},
  author={Fujimoto, Scott and Meger, David and Precup, Doina},
  booktitle={International conference on machine learning},
  pages={2052--2062},
  year={2019},
  organization={PMLR}
}

@article{foster2024behavior,
  title={Is behavior cloning all you need? understanding horizon in imitation learning},
  author={Foster, Dylan J and Block, Adam and Misra, Dipendra},
  journal={arXiv preprint arXiv:2407.15007},
  year={2024}
}

@article{zeng2023demonstrations,
  title={When demonstrations meet generative world models: A maximum likelihood framework for offline inverse reinforcement learning},
  author={Zeng, Siliang and Li, Chenliang and Garcia, Alfredo and Hong, Mingyi},
  journal={Advances in Neural Information Processing Systems},
  volume={36},
  pages={65531--65565},
  year={2023}
}

@article{zhao2024mobility,
  title={Mobility-aware computation offloading for AR tasks over terahertz wireless networks: An offline reinforcement learning approach},
  author={Zhao, Shuyue and Jing, Wenpeng and Yu, F Richard and Wen, Xiangming and Lu, Zhaoming},
  journal={IEEE Transactions on Vehicular Technology},
  volume={73},
  number={12},
  pages={19111--19124},
  year={2024},
  publisher={IEEE}
}

@ARTICLE{Nguyen2021AINetwork,
  author={Nguyen, Dinh C. and Cheng, Peng and Ding, Ming and Lopez-Perez, David and Pathirana, Pubudu N. and Li, Jun and Seneviratne, Aruna and Li, Yonghui and Poor, H. Vincent},
  journal={IEEE Communications Surveys and Tutorials}, 
  title={Enabling AI in Future Wireless Networks: A Data Life Cycle Perspective}, 
  year={2021},
  volume={23},
  number={1},
  pages={553-595}}

@INPROCEEDINGS{Tan2022offloading,
  author={Tan, Jing and Khalili, Ramin and Karl, Holger and Hecker, Artur},
  booktitle={IEEE Conference on Computer Communications}, 
  title={Multi-Agent Distributed Reinforcement Learning for Making Decentralized Offloading Decisions}, 
  year={2022},
  volume={},
  number={},
  pages={2098-2107}}

@ARTICLE{Yang2025OnlineRL,
  author={Yang, Ning and Chen, Shuo and Zhang, Haijun and Berry, Randall},
  journal={IEEE Communications Surveys and Tutorials}, 
  title={Beyond the Edge: An Advanced Exploration of Reinforcement Learning for Mobile Edge Computing, Its Applications, and Future Research Trajectories}, 
  year={2025},
  volume={27},
  number={1},
  pages={546-594}}

@ARTICLE{Yang2024OnlineIss,
  author={Yang, Kun and Shi, Chengshuai and Shen, Cong and Yang, Jing and Yeh, Shu-Ping and Sydir, Jaroslaw J.},
  journal={IEEE Transactions on Wireless Communications}, 
  title={Offline Reinforcement Learning for Wireless Network Optimization With Mixture Datasets}, 
  year={2024},
  volume={23},
  number={10},
  pages={12703-12716}}

@ARTICLE{Fig2024OfflineSurv,
  author={Figueiredo Prudencio, Rafael and Maximo, Marcos R. O. A. and Colombini, Esther Luna},
  journal={IEEE Transactions on Neural Networks and Learning Systems}, 
  title={A Survey on Offline Reinforcement Learning: Taxonomy, Review, and Open Problems}, 
  year={2024},
  volume={35},
  number={8},
  }

@misc{Adroit2018,
      title={Learning Complex Dexterous Manipulation with Deep Reinforcement Learning and Demonstrations}, 
      author={Aravind Rajeswaran and Vikash Kumar and Abhishek Gupta and Giulia Vezzani and John Schulman and Emanuel Todorov and Sergey Levine},
      year={2018},
      eprint={1709.10087},
      archivePrefix={arXiv}, 
      primaryClass={cs.LG}
}

@inproceedings{todorov2012mujoco,
  title={MuJoCo: A physics engine for model-based control},
  author={Todorov, Emanuel and Erez, Tom and Tassa, Yuval},
  booktitle={2012 IEEE/RSJ International Conference on Intelligent Robots and Systems},
  pages={5026--5033},
  year={2012},
  organization={IEEE},
  doi={10.1109/IROS.2012.6386109}
}

@misc{fu2020d4rl,
    title={D4RL: Datasets for Deep Data-Driven Reinforcement Learning},
    author={Justin Fu and Aviral Kumar and Ofir Nachum and George Tucker and Sergey Levine},
    year={2020},
    eprint={2004.07219},
    archivePrefix={arXiv},
    primaryClass={cs.LG}
}

@misc{YOLOV11,
      title={YOLOv11: An Overview of the Key Architectural Enhancements}, 
      author={Rahima Khanam and Muhammad Hussain},
      year={2024},
      eprint={2410.17725},
      archivePrefix={arXiv},
      primaryClass={cs.CV}
}

@article{balazadeh2024sequential,
  title={Sequential Decision Making with Expert Demonstrations under Unobserved Heterogeneity},
  author={Balazadeh, Vahid and Chidambaram, Keertana and Nguyen, Viet and Krishnan, Rahul G and Syrgkanis, Vasilis},
  journal={Advances in Neural Information Processing Systems},
  volume={37},
  pages={65476--65498},
  year={2024}
}

@inproceedings{
pace2023delphic,
title={Delphic Offline Reinforcement Learning under Nonidentifiable Hidden Confounding},
author={Aliz{\'e}e Pace and Hugo Y{\`e}che and Bernhard Sch{\"o}lkopf and Gunnar Ratsch and Guy Tennenholtz},
booktitle={The Twelfth International Conference on Learning Representations},
year={2024},
}

@inproceedings{qiao2023privacy,
author = {Qiao, Dan and Wang, Yu-Xiang},
title = {Offline reinforcement learning with differential privacy},
year = {2023},
booktitle = {Proceedings of the 37th International Conference on Neural Information Processing Systems},
}

@inproceedings{
zhao2025beyondexpert,
title={Beyond-Expert Performance with Limited Demonstrations: Efficient Imitation Learning with Double Exploration},
author={Heyang Zhao and Xingrui Yu and David Mark Bossens and Ivor Tsang and Quanquan Gu},
booktitle={The Thirteenth International Conference on Learning Representations},
year={2025},
url={https://openreview.net/forum?id=FviefuxmeW}
}

@inproceedings{wang2026trajectory,
title={Trajectory Generation with Conservative Value Guidance for Offline Reinforcement Learning},
author={Tieru Wang and Kunbao Wu and Guoshun Nan},
booktitle={The Fourteenth International Conference on Learning Representations},
year={2026},
}

@inproceedings{yang2024advancing,
  title={Advancing RAN slicing with offline reinforcement learning},
  author={Yang, Kun and Yeh, Shu-Ping and Zhang, Menglei and Sydir, Jerry and Yang, Jing and Shen, Cong},
  booktitle={2024 IEEE international symposium on dynamic spectrum access networks},
  pages={331--338},
  year={2024},
  organization={IEEE}
}

@inproceedings{wang2025beyond,
  title={Beyond Static Populations: Efficient Delay-Constrained Scheduling for Dynamic Users via Deep Reinforcement Learning},
  author={Wang, Xun and Li, Zhuoran and Huang, Longbo},
  booktitle={Proceedings of the Twenty-sixth International Symposium on Theory, Algorithmic Foundations, and Protocol Design for Mobile Networks and Mobile Computing},
  pages={151--160},
  year={2025}
}

@article{liu2023optimizing,
  title={Optimizing age of information in wireless uplink networks with partial observations},
  author={Liu, Jingwei and Zhang, Rui and Gong, Aoyu and Chen, He},
  journal={IEEE Transactions on Communications},
  volume={71},
  number={7},
  pages={4105--4118},
  year={2023},
  publisher={IEEE}
}

@article{zangooei2023reinforcement,
  title={Reinforcement learning for radio resource management in RAN slicing: A survey},
  author={Zangooei, Mohammad and Saha, Niloy and Golkarifard, Morteza and Boutaba, Raouf},
  journal={IEEE Communications Magazine},
  volume={61},
  number={2},
  pages={118--124},
  year={2023},
  publisher={IEEE}
}

@article{ammar2024handoffs,
  title={Handoffs in user-centric cell-free MIMO networks: A POMDP framework},
  author={Ammar, Hussein A and Adve, Raviraj and Shahbazpanahi, Shahram and Boudreau, Gary and Srinivas, Kothapalli Venkata},
  journal={IEEE Transactions on Wireless Communications},
  volume={23},
  number={8},
  pages={10319--10335},
  year={2024},
  publisher={IEEE}
}

@inproceedings{oliveira2021generating,
  title={Generating synthetic datasets for mobile wireless networks with sumo},
  author={Oliveira, Afonso and Vaz{\~a}o, Teresa},
  booktitle={Proceedings of the 19th ACM international symposium on mobility management and wireless access},
  pages={33--42},
  year={2021}
}

@misc{kostrikov2021implicit,
      title={Offline Reinforcement Learning with Implicit Q-Learning}, 
      author={Ilya Kostrikov and Ashvin Nair and Sergey Levine},
      year={2021},
      eprint={2110.06169},
      archivePrefix={arXiv},
      primaryClass={cs.LG},
}

@article{kiruthiga2025evolution,
  title={Evolution of next generation networks and its contribution towards industry 5.0},
  author={Kiruthiga Devi, M and Padma Priya, M},
  journal={Resource management in advanced wireless networks},
  pages={45--80},
  year={2025},
  publisher={Wiley Online Library}
}

@article{fu2024robot,
  title={Robot policy learning with temporal optimal transport reward},
  author={Fu, Yuwei and Zhang, Haichao and Wu, Di and Xu, Wei and Boulet, Benoit},
  journal={Advances in Neural Information Processing Systems},
  volume={37},
  pages={122078--122103},
  year={2024}
}

@article{
  chaudhary2025from,
  title={From Novelty to Imitation: Self-Distilled Rewards for Offline Reinforcement Learning},
  author={Gaurav Chaudhary and Laxmidhar Behera},
  journal={Transactions on Machine Learning Research},
  issn={2835-8856},
  year={2025},
}

@InProceedings{pmlr-v162-ni22a,
  title = 	 {Recurrent Model-Free {RL} Can Be a Strong Baseline for Many {POMDP}s},
  author =       {Ni, Tianwei and Eysenbach, Benjamin and Salakhutdinov, Ruslan},
  booktitle = 	 {International Conference on Machine Learning},
  pages = 	 {16691--16723},
  year = 	 {2022},
  volume = 	 {162},
  month = 	 {17--23 Jul},
  publisher =    {PMLR},
}

@inproceedings{
han2020variational,
title={Variational Recurrent Models for Solving Partially Observable Control Tasks},
author={Dongqi Han and Kenji Doya and Jun Tani},
booktitle={International Conference on Learning Representations},
year={2020},
}

\end{document}